\documentclass[aps,floatfix,amsmath,nofootinbib,amssymb,superscriptaddress]{revtex4}

\usepackage{overpic}
\usepackage{amssymb}
\usepackage{indentfirst}
\usepackage{feynmf}   
\usepackage{slashed}  
\usepackage{cases}
\usepackage{color}
\usepackage{multirow}
\usepackage{epstopdf}
\usepackage{graphicx,color,bm}
\usepackage{epstopdf}

\usepackage[colorlinks,
            citecolor=blue,
            anchorcolor=red,
            menucolor=red,
            linkcolor=red,
            filecolor=red,
            runcolor=red,
            urlcolor=blue,
            frenchlinks=red]{hyperref}

\begin{document}

\title{The $\sin2\phi_h$ azimuthal asymmetry of pion production in SIDIS within TMD factorization}
\author{Hui Li}
\affiliation{School of Physics, Southeast University, Nanjing 211189, China}
\author{Zhun Lu}
\email{zhunlu@seu.edu.cn}
\affiliation{School of Physics, Southeast University, Nanjing 211189, China}

\begin{abstract}

We study the $\sin2\phi_h$ azimuthal asymmetry of charged and neutral pion productions off the longitudinally polarized nucleon targets in semi-inclusive deeply inelastic scattering (SIDIS) process within the transverse momentum dependent~(TMD) factorization.
The asymmetry is contributed by the convolution of the TMD distribution $h_{1L}^{\perp}$ and the Collins fragmentation function $H_{1}^\perp$.
We adopt the Wandzura-Wilczek-type~(WW-type) approximation for $h_{1L}^{\perp}$ and two different parametrizations on the nonperturbative part of TMD evolution for $h_{1L}^{\perp}$ and $H_{1}^\perp$.
We numerically estimate the asymmetry in the kinematical configurations available at HERMES, CLAS and CLAS12.
It is shown that our theoretical calculation can describe the HERMES and CLAS data, except the asymmetry of the $\pi^-$ production off proton target at HERMES and CLAS.
We also find that different choices of the nonperturbative part of TMD evolution lead qualitatively similar results.
\end{abstract}

\maketitle

\section{Introduction}

Investigating the internal structure of the nucleon is still a frontier of hadronic physics.
An important catalog of observables is the transverse momentum dependent distributions (TMDs) since they encode three-dimensional information of the nucleon in the momentum space structures~\cite{Sivers:1989cc,Sivers:1990fh,Anselmino:1994tv,Anselmino:1998yz,Brodsky:2002cx,Brodsky:2002rv,
Collins:2002kn,Ji:2002aa,Belitsky:2002sm,Boer:2003cm,Ji:2006ub,Ji:2006vf,Ji:2006br,Bacchetta:2008xw}, which is much richer than that of the collinear parton distribution functions. At leading twist, there are eight TMD PDFs appearing in the decomposition of the quark-quark correlator of the nucleon~\cite{Mulders:1995dh,Boer:1997nt,Bacchetta:2006tn}. Each of these TMDs, depending on the longitudinal momentum fraction $x$ and the transverse momentum $\bm{p}_T$, represents a special spin and partonic structure of the nucleon. The essential tools to explore TMDs are the spin and azimuthal asymmetries in various polarized or unpolarized processes involving at least two hadrons, such as SIDIS~\cite{HERMES:2004mhh,HERMES:2009lmz,HERMES:2010mmo,COMPASS:2005csq,COMPASS:2006mkl,COMPASS:2010hbb,Mkrtchyan:2007sr,CLAS:2008nzy,
EuropeanMuon:1986ulc,ZEUS:2000esx,Kafer:2008ud,Bressan:2009eu} and Drell-Yan (DY)~\cite{NA10:1986fgk,NA10:1987sqk,NuSea:2007ult,NuSea:2008ndg} processes.

Among the leading-twist TMDs, the distribution $h_{1L}^{\perp}(x,\bm{p}_T^2)$~\cite{Kotzinian:1994dv} is of particular importance. It describes the probability of finding a transversely polarized quark but inside a longitudinally polarized nucleon. Therefore, it is also called as the worm-gear L distribution or longi-transversity.
Since $h_{1L}^{\perp}$ is chiral odd, it has to be coupled to another chiral-odd function to manifest its effects in semi-inclusive processes. In SIDIS, this can be achieved via a $\sin2\phi_{\pi}$ azimuthal asymmetry~\cite{Kotzinian:1994dv,Kotzinian:1997wt} when $h_{1L}^{\perp}$ is combined with the chiral-odd Collins function $H_1^\perp$~\cite{Collins:1992kk}.
The early work on the $\sin2\phi_{\pi}$ asymmetry in the longitudinally polarized SIDIS process have been performed in Refs.~\cite{Airapetian:1999tv,Airapetian:2002mf,Avakian:2010ae,Lu:2011pt,
Zhu:2011zza,Boffi:2009sh,Ma:2000ip,Ma:2001ie}, showing that the asymmetry is around several percents.

In this paper, we perform a detailed phenomenological analysis of the $\sin2\phi_h$ azimuthal asymmetry of the charged and neutral pion productions off the longitudinally polarized nucleon target in SIDIS within TMD factorization~\cite{Collins:1981uk,Collins:1984kg,Ji:2004wu,Ji:2004xq,Collins:2011zzd}.
We consider the kinematical region accessible at HERMES, CLAS and CLAS12
And compare the results with the available data at HERMES (deuteron and proton target)~\cite{Avakian:1999rr,Avakian:2007mv,Airapetian:1999tv, Airapetian:2001eg,Airapetian:2002mf} and CLAS (proton target)~\cite{Jawalkar:2017ube,Avakian:2010ae} data.
We apply the TMD factorization to estimate the spin-dependent cross section in
$l+N^\rightarrow\rightarrow l+\pi+X$ as well as the unpolarized cross section.
The asymmetry can be expressed as the ratio of the two cross sections.
In the last two decades, TMD factorization has become a powerful tool for studying the three-dimensional structure of the nucleon and has been widely applied in various high energy processes~\cite{Boer:2008fr,Arnold:2008kf,Aybat:2011zv,Collins:2011zzd,Collins:2012uy,Echevarria:2012pw,
Echevarria:2012js,Pitonyak:2013dsu,Echevarria:2014xaa,Kang:2015msa,Bacchetta:2017gcc,Wang:2017zym,Wang:2018pmx,
Li:2019uhj}.
The TMD factorization theorem allows the differential cross section in the small transverse momentum region $P_{\pi T}/z \ll Q$~($P_{\pi T}$ is the transverse momentum of the $\pi$ production and $Q$ is the virtuality of the photon) to be expressed as a convolution of two contributions: one corresponds to the hard scattering factors at short distance; the other accounts for the coherent long-distance interactions, and is described in terms of the well-defined TMDs.
In our case, the $\sin2\phi_{\pi}$ asymmetry is contributed by the convolution of longitudinal transversity PDF $h_{1L}^{\perp}$, Collins FF $H_1^\perp$ and the hard scattering factors.
The TMD formalism also encodes the evolution information of TMDs, of which the energy evolution (or the scale dependence) are governed by the so-called Collins-Soper equation~\cite{Collins:1981uk,Collins:1984kg,Collins:2011zzd,Idilbi:2004vb}.
The solution of the equation is usually expressed as an exponential form of the Sudakov-like form factor~\cite{Collins:1984kg,Collins:2011zzd,Aybat:2011zv,Collins:1999dz} which indicates the change of TMDs from a initial scale to another scale.
The Sudakov-like form factor can be separated to two parts. One is the perturbative part, which can be calculated perturbatively and is the same for different TMDs; the other is the nonperturbative part, which can not be calculated directly and is usually obtained by phenomenological extraction from experimental data.
In the literature, several nonperturbative parts of the Sudakov form factor for the TMDs have been extracted from experimental data~\cite{Collins:1984kg,Davies:1984sp,Ellis:1997sc,Landry:2002ix,Konychev:2005iy,
Collins:2011zzd,Aybat:2011zv,Aybat:2011ge,Kang:2011mr,Echevarria:2012js,Su:2014wpa,
Echevarria:2014xaa,Echevarria:2014rua,Bacchetta:2017gcc}.
In this work, we will consider the TMD evolution effect of TMDs with two parameterizations on the nonperturbative part~\cite{Echevarria:2014xaa,Bacchetta:2017gcc} to estimate the asymmetry for comparison.

The remaining content of the paper is organized as follows.
In Sec.~\ref{Sec.formalism}, we investigate the evolution effect for the TMDs.
Particularly, we discuss the parametrization of the nonperturbative Sudakov form factors associated with the studied TMDs in details.
In Sec.~\ref{Sec.evolution}, we present the formalism of the $\sin 2\phi_h$ asymmetry in $l N^\rightarrow\to l\pi X$  within the TMD factorization.
In Sec.~\ref{Sec.numerical}, we numerically estimate the asymmetry $A_{UL}^{\sin2\phi_h}$ in the $l N^\rightarrow\to l\pi X$ process at the kinematical region of HERMES, CLAS, CLAS12 with two different choices on the nonperturbative part associated with TMD evolution effect.
Finally, We summarize the paper in Sec.~\ref{Sec.conclusion}.

\section{The evolution of TMDs}
\label{Sec.evolution}

In this section, we review the evolution formalism of the unpolarized PDF $f_1$, the longitudinal transversity PDF $h_{1L}^{\perp}$ of the nucleon as well as the unpolarized FF $D_1$, the Collins FF $H_{1}^\perp$ of the pion, within the TMD factorization.

TMD evolution is usually performed in coordinate $b$-space, since the cross section can be expressed as the simpler product in $b$ space than in the transverse momentum $k_{\perp}$ space ($b$-space is conjugated to $k_{\perp}$-space via Fourier transformation)~\cite{Collins:1984kg,Collins:2011zzd}.
In the TMD factorization based on different schemes (such as the CS-81~\cite{Collins:1981uk}, JMY~\cite{Ji:2004xq,Ji:2004wu} and Collins-11 schemes~\cite{Collins:2011zzd}), the TMDs $\tilde{F}(x,b;\mu,\zeta_F)$  and $\tilde{D}(z,b;\mu,\zeta_D)$) depend on two energy scales~\cite{Collins:1981uk,Collins:1984kg,Collins:2011zzd,Aybat:2011zv,Aybat:2011ge,Echevarria:2012pw}. One is the renormalization scale $\mu$, the other is the energy scale $\zeta_F$ (or $\zeta_D$) serving as a cutoff to regularize the light-cone singularity in the operator definition of the TMDs.
The two energy dependencies are encoded in different evolution equations.
For the $\zeta_F$ (or $\zeta_D$) dependence of the TMD PDFs (or FFs), it is determined by the Collins-Soper~(CS) equation ~\cite{Collins:1981uk} (in this paper $b= |\bm b_\perp|$ and the tilde terms represent the ones in $b$ space):
\begin{align}
\frac{\partial\ \mathrm{ln} \tilde{F}(x,b;\mu,\zeta_F)}{\partial\ \sqrt{\zeta_F}}=\frac{\partial\ \mathrm{ln} \tilde{D}(z,b;\mu,\zeta_D)}{\partial\ \sqrt{\zeta_D}}=\tilde{K}(b;\mu),\label{eq:K}
\end{align}
with $\tilde{K}$ being the CS evolution kernel which can be computed perturbatively for small values of $b$(up to order $\alpha_s$):
\begin{align}
\tilde{K}(b;\mu)=-\frac{\alpha_s C_F}{\pi} \left[\ln(\mu^2 b^2)-\ln 4 +2\gamma_E\right]+\mathcal{O}(\alpha_s^2),
\end{align}
and $\gamma_E\approx0.577$ is the Euler's constant~\cite{Collins:1981uk}.

For the $\mu$ dependence of the TMDs, it is derived from the renormalization group equation
\begin{align}
&\frac{d\ \tilde{K}}{d\ \mathrm{ln}\mu}=-\gamma_K(\alpha_s(\mu)),\label{eq:evo1}\\
&\frac{d\ \mathrm{ln} \tilde{F}(x,b;\mu,\zeta_F)}
{d\ \mathrm{ln}\mu}=\gamma_F(\alpha_s(\mu);{\frac{\zeta^2_F}{\mu^2}}),\label{eq:evo2}\\
&\frac{d\ \mathrm{ln} \tilde{D}(z,b;\mu,\zeta_D)}
{d\ \mathrm{ln}\mu}=\gamma_D(\alpha_s(\mu);{\frac{\zeta^2_D}{\mu^2}}),\label{eq:evo3}
\end{align}
where $\gamma_K$, $\gamma_F$ and $\gamma_D$ are the anomalous dimensions of $\tilde{K}$, $\tilde{F}$ and $\tilde{D}$, respectively,
\begin{align}
\gamma_K & =2{\alpha_s C_F\over \pi}+\mathcal{O}(\alpha_s^2), \\
\gamma_D &=\gamma_F = \alpha_s {C_F \over \pi}\left({3\over 2} - \ln\left(\zeta_F\over \mu^2\right)\right)+\mathcal{O}(\alpha_s^2).
\end{align}
Solving Eqs.~(\ref{eq:evo1}-\ref{eq:evo3}), one can obtain the general solution for the energy dependence of $\tilde{F}$ (or $\tilde{D}$)~\cite{Idilbi:2004vb,Collins:1981uk,Collins:1984kg,Collins:2011zzd,Ji:2004wu,Collins:2014jpa,Ji:2004xq}:
\begin{align}
\tilde{F}(x,b;Q)=\mathcal{F}\times e^{-S(Q,b)}\times \tilde{F}(x,b;\mu_i),
\label{eq:Sudakov factor F}\\
\tilde{D}(z,b;Q)=\mathcal{D}\times e^{-S(Q,b)}\times \tilde{D}(z,b;\mu_i),
\label{eq:Sudakov factor D}
\end{align}
where $\mathcal{F}$ and $\mathcal{D}$ are the factors related to the hard scattering, $S(Q,b)$ is the Sudakov form factor. Hereafter, we will set $\mu=\sqrt{\zeta_F}= \sqrt{\zeta_D}=Q$, and express $\tilde{F}(x,b;\mu=Q,\zeta_F=Q^2)$ (or $\tilde{D}(z,b;\mu=Q,\zeta_D=Q^2)$) as $\tilde{F}(x,b;Q)$ (or $\tilde{D}(z,b;Q)$) for simplicity.
Eq.~(\ref{eq:Sudakov factor F}) (or Eq.~(\ref{eq:Sudakov factor D})) demonstrates that the distribution $\tilde{F}$ (or $\tilde{D}$) at an arbitrary scale $Q$ can be determined by the same distribution at an initial scale $\mu_i$ through the evolution encoded by the exponential form $\mathrm{exp}(-S(Q,b))$.

To be more specific, the exponential exp($-S(Q,b)$) has the following explicit form  (taking the one for $\tilde{F}$ as an example)
\begin{align}
\label{eq:exp}
\exp(-S(Q,b)) &=  \exp \left\{ \ln \frac{Q}{\mu} \tilde{K}(b_{\ast};\mu) +
\int_{\mu_i}^\mu \frac{d \bar{\mu}}{\bar{\mu}} \left[ \gamma_F(g(\bar{\mu});1)
- \ln \left(\frac{Q}{\bar{\mu}}\right) \gamma_K(g(\bar{\mu})) \right]\right\}\nonumber\\
&\times
\exp \left\{ g_{j/P}(x,b) + g_K(b) \ln \frac{Q}{Q_0} \right\}.
\end{align}
The exponential in the first line of Eq.~(\ref{eq:exp}) comes from the solutions of Eqs.~(\ref{eq:K}), (\ref{eq:evo1}) and (\ref{eq:evo2}) in the perturbative region (the small $b$ region $ b \ll 1/ \Lambda$). It contains $\tilde{K}(b_{\ast};\mu)$, the CS evolution kernel in the small $b$ region, and the anomalous dimension $\gamma_F$, $\gamma_K$.
However, in the nonperturbative region (large $b$ region), the evolution kernel $\tilde K(b;\mu)$ is not calculable.
In order to access the contribution in the large $b$ region, the exponential in the second line of Eq.~(\ref{eq:exp}) is introduced.
Here, the function $g_{j/P}(x,b)$ parametrizes the non-perturbative large-$b$ behavior that is intrinsic to the proton target, while the universal function $g_K$ parametrizes the non-perturbative large-$b$ behavior of the evolution kernel $\tilde{K}(b;\mu)$.

To allow a smooth transition of $b$ from perturbative region to nonperturbative region as well as to avoid hitting on the Landau pole, one can set a parameter $b_{\mathrm{max}}$ to be the boundary between the two regions.
The typical value of $b_{\mathrm{max}}$ is chosen around $1\ \mathrm{GeV}^{-1}$ to guarantee that $b_{\ast}$ is always in the perturbative region.
A $b$-dependent function $b_\ast(b)$ may be also introduced to have the property $b_\ast\approx b$ at small $b$ value and $b_{\ast}\approx b_{\mathrm{max}}$ at large $b$ value.
There are several different choices on the form of $b_\ast(b)$ in the literature~\cite{Collins:1984kg,Collins:2016hqq,Bacchetta:2017gcc}. A frequently used one is the Collins-Soper-Sterman (CSS) prescription~\cite{Collins:1984kg}:
\begin{align}
\label{eq:b*}
b_\ast=b/\sqrt{1+b^2/b_{\rm max}^2}  \ ,~b_{\rm max}<1/\Lambda_\mathrm{QCD}.
\end{align}

Combining the perturbative part and the nonperturbative part, one has the complete result for the Sudakov form factor appearing in Eqs.~(\ref{eq:Sudakov factor F}) and (\ref{eq:Sudakov factor D}):
\begin{equation}
\label{eq:S}
S(Q,b)=S_{\mathrm{P}}(Q,b_\ast)+S_{\mathrm{NP}}(Q,b).
\end{equation}
with the boundary of the two parts set by the $b_\mathrm {max}$.
The perturbative part $S_{\mathrm{P}}(Q,b_\ast)$ has been studied~\cite{Echevarria:2014xaa,Kang:2011mr,Aybat:2011ge,Echevarria:2012pw,Echevarria:2014rua} in details and has the following form:
\begin{equation}
\label{eq:Spert}
S_{\mathrm{P}}(Q,b_\ast)=\int^{Q^2}_{\mu_b^2}\frac{d\bar{\mu}^2}{\bar{\mu}^2}\left[A(\alpha_s(\bar{\mu}))
\mathrm{ln}\frac{Q^2}{\bar{\mu}^2}+B(\alpha_s(\bar{\mu}))\right],
\end{equation}
which is the same for different kinds of PDFs and FFs, namely, $S_{P}$ is spin-independent. In addition, the coefficients $A$ and $B$ in Eq.(\ref{eq:Spert}) can be expanded as the series of $\alpha_s/{\pi}$:
\begin{align}
A=\sum_{n=1}^{\infty}A^{(n)}(\frac{\alpha_s}{\pi})^n,\\
B=\sum_{n=1}^{\infty}B^{(n)}(\frac{\alpha_s}{\pi})^n.
\end{align}
In this work, we will take $A^{(n)}$ up to $A^{(2)}$ and $B^{(n)}$ up to $B^{(1)}$ in the accuracy of next-to-leading-logarithmic (NLL) order~\cite{Collins:1984kg,Landry:2002ix,Qiu:2000ga,Kang:2011mr,Aybat:2011zv,Echevarria:2012pw} :
\begin{align}
A^{(1)}&=C_F,\\
A^{(2)}&=\frac{C_F}{2}\left[C_A\left(\frac{67}{18}-\frac{\pi^2}{6}\right)-\frac{10}{9}T_Rn_f\right],\\
B^{(1)}&=-\frac{3}{2}C_F,
\end{align}
with $C_{F}=4/3$, $C_{A}=3$ and $T_{R}=1/2$.

The non-perturbative part $S_\mathrm{NP}$ in Eq.(\ref{eq:S}) can not be calculated perturbatively, it is usually parameterized and extracted from experimental data.
There are several different parametrizations on $S_\mathrm{NP}$ in the literature~\cite{Collins:1984kg,Davies:1984sp,Ellis:1997sc,Landry:2002ix,Konychev:2005iy,
Collins:2011zzd,Aybat:2011zv,Aybat:2011ge,Kang:2011mr,Echevarria:2012js,Su:2014wpa,
Echevarria:2014xaa,Echevarria:2014rua,Bacchetta:2017gcc,Collins:2014jpa,Nadolsky:1999kb,Aidala:2014hva},
we will discuss two of them in details to investigate the impact of the different evolution formalisms on the asymmetry.

One of the parametrizations applied in this study is the Echevarria-Idilbi-Kang-Vitev (EIKV parametrization) non-perturbative Sudakov $S_\mathrm{NP}$ for the unpolarized TMD PDFs (or TMD FFs), which has the following form~\cite{Echevarria:2014xaa}:
\begin{equation}
S^\mathrm{pdf}_\mathrm{NP}(b,Q)=b^2(g_1^\mathrm{pdf}+\frac{g_2}{2}\ln{\frac{Q}{Q_0}}), \label{eq:eikv1}
\end{equation}
\begin{equation}
S^\mathrm{ff}_\mathrm{NP}(b,Q)=b^2(g_1^\mathrm{ff}+\frac{g_2}{2}\ln{\frac{Q}{Q_0}}).\label{eq:eikv2}
\end{equation}
Here, $g_2$ includes the information on the large $b$ behavior of the evolution kernel $\tilde{K}$ ($g_K(b)=g_2 b^2$).
This function is universal for different types of TMDs and does not depend on the particular process, which is an important prediction of QCD factorization theorems involving TMDs~\cite{Aybat:2011zv,Collins:2011zzd,Echevarria:2014xaa,Kang:2015msa}.
$g_1$ contains information on the intrinsic nonperturbative transverse motion of bound partons.
It could depend on the type of TMDs, and can be interpreted as the intrinsic transverse momentum width for the relevant TMDs at the initial scale $Q_0$~\cite{Aybat:2011zv,Qiu:2000ga,Qiu:2000hf,Anselmino:2012aa,Su:2014wpa}.
Furthermore, $g_1^\mathrm{pdf}$ and $g_1^\mathrm{ff}$ are parameterized as:
\begin{equation}
g_1^\mathrm{pdf}=\frac{\langle k_T^2\rangle_{Q_0}}{4},
\end{equation}
\begin{equation}
g_1^\mathrm{ff}=\frac{\langle p_T^2\rangle_{Q_0}}{4z^2},
\end{equation}
where $\langle k_T^2\rangle_{Q_0}$ and $\langle p_T^2\rangle_{Q_0}$ are the averaged intrinsic transverse momenta squared for TMD PDFs and FFs at the initial scale $Q_0$, respectively.
In Ref.~\cite{Echevarria:2014xaa} the authors tuned the current extracted ranges of three parameters $\langle k_T^2\rangle_{Q_0}, \langle p_T^2\rangle_{Q_0}$ and $g_2$ with $Q_0=\sqrt{2.4}\ \textrm{GeV}$ in
Refs.~\cite{Anselmino2005,Collins:2005ie,Schweitzer:2010tt} and further found that the following fixed values of parameters can reasonably describe the SIDIS data together with the Drell-Yan lepton pair and $W/Z$ boson production data:
\begin{equation}
\langle k_T^2\rangle_{Q_0}=0.38\ \textrm{GeV}^2, ~~~~~~\langle p_T^2\rangle_{Q_0}=0.19\ \textrm{GeV}^2, ~~~~~~g_2=0.16\ \textrm{GeV}^2,~~~~~b_\mathrm{max}=1.5\ \textrm{GeV}^{-1}.
\end{equation}

Besides the Sudakov form factor in Eqs.(\ref{eq:Sudakov factor F}) and (\ref{eq:Sudakov factor D}), another important element in Eqs.(\ref{eq:Sudakov factor F}) and (\ref{eq:Sudakov factor D}) is the TMDs at a fixed scale $\mu$. In the small $b$ region, $\tilde{F}(x,b;\mu)$ and $\tilde{D}(x,b;\mu)$ at a fixed scale $\mu$ can be expressed as the convolution of the perturbatively calculable coefficients $C$ and the corresponding collinear counterparts $F_{i/H}(\xi,\mu)$ (or $D_{H/j}(\xi,\mu)$),
\begin{align}
\tilde{F}(x,b;\mu)=\sum_i \int_{x}^1\frac{d\xi}{\xi} C_{q\leftarrow i}(x/\xi,b;\mu)F_{i/H}(\xi,\mu),
\label{eq:small_b_F}\\
\tilde{D}(z,b;\mu)=\sum_j \int_{z}^1\frac{d\xi}{\xi} C_{j\leftarrow q}(z/\xi,b;\mu)D_{H/j}(\xi,\mu),
\label{eq:small_b_D}
\end{align}
Here, $\mu$ is a dynamic scale related to $b_\ast$ by $\mu=c/b_\ast$ , with $c=2e^{-\gamma_E}$ and $\gamma_E\approx0.577$ being the Euler's constant~\cite{Collins:1981uk}, $C_{q\leftarrow i}(x/\xi,b;\mu)=\sum_{n=0}^{\infty}C_{q\leftarrow i}^{(n)}(\alpha_s/\pi)^n$ and $C_{j\leftarrow q}(z/\xi,b;\mu)=\sum_{n=0}^{\infty}C_{j\leftarrow q}^{(n)}(\alpha_s/\pi)^n$ are the perturatively calculable coefficient function.

After solving the evolution equations and incorporating the Sudakov form factor, the scale-dependent TMDs in $b$ space can be rewritten as
\begin{align}
\label{eq:tildeF}
\tilde{F}_{q/H}(x,b;Q)=e^{-\frac{1}{2}S_{\mathrm{P}}(Q,b_\ast)-S^{F_{q/H}}_{\mathrm{NP}}(Q,b)}F_{q/H}(x,\mu),\\
\tilde{D}_{H/q}(z,b;Q)=e^{-\frac{1}{2}S_{\mathrm{P}}(Q,b_\ast)-S^{D_{H/q}}_{\mathrm{NP}}(Q,b)}D_{H/q}(z,\mu)
\end{align}
The factor of $\frac{1}{2}$ in front of $S_{\mathrm{P}}$ comes from the fact that $S_{\mathrm{P}}$ is equally distributed to the initial-state quark and the final-state quark~\cite{Prokudin:2015ysa}.
In this work, we adopt the leading order (LO) results for the hard coefficients $C$, $\mathcal{F}$ and $\mathcal{D}$ for $f_1$, $h_{1L}^{\perp}$, $D_1$ and $H_{1}^\perp$, i.e. $C_{q\leftarrow i}^{(0)}=\delta_{iq}\delta(1-x)$, $C_{j\leftarrow q}^{(0)}=\delta_{qj}\delta(1-z)$, $\mathcal{F}=1$ and $\mathcal{D}=1$.

With all the ingredients above, we can obtain the unpolarized PDFs of nucleon $\tilde{f}_1^{q/p}$ (or FFs of pion $\tilde{D}_1^{\pi/q}$) in $b$ space
\begin{align}
\tilde{f}_1^{q/p}(x,b;Q) &=e^{-\frac{1}{2}S_{\mathrm{P}}(Q,b_\ast)-S^{pdf}_{\mathrm{NP}}(Q,b)}
 f_1^{q/p}(x,\mu),\label{eq:f_b}\\
 \tilde{D}_1^{\pi/q}(z,b;Q) &=e^{-\frac{1}{2}S_{\mathrm{P}}(Q,b_\ast)-S^{ff}_{\mathrm{NP}}(Q,b)}
 D_1^{\pi/q}(z,\mu)\label{eq:D_b}.
\end{align}

By performing the Fourier transformation, we can obtain the unpolarized PDFs of nucleon $f_1^{q/N}$ (or FFs of pion $D_1^{\pi/q}$) in the transverse momentum space
\begin{align}
f_1^{q/N}(x,\bm{p}_T;Q)&=\int_0^\infty\frac{db b}{2\pi}J_0(|\bm p_T| b)e^{-\frac{1}{2}S_{P}(Q,b_\ast)-S^\mathrm{pdf}_{\mathrm{NP}}(Q,b)} f_1^{q/p}(x,\mu),\label{eq:f1}\\
D_1^{\pi/q}(z,\bm{K}_\perp;Q)&=\int_0^\infty\frac{db b}{2\pi}J_0(|\bm K_\perp|  b/z)e^{-\frac{1}{2}S_{P}(Q,b_\ast)-S^\mathrm{ff}_{\mathrm{NP}}(Q,b)} D_1^{\Lambda/q}(z,\mu).\label{eq:D1pi}
\end{align}

According to Eqs.(\ref{eq:small_b_F}) and (\ref{eq:small_b_D}), in the small $b$ region, we can also express the longitudinal transversity PDF of nucleon target $\tilde{h}_{1L}^{\perp}$ and the Collins FF of pion production $\tilde{H}_{1}^{\perp}$ at a fixed energy scale $\mu$ in terms of the perturbatively calculable coefficients and the corresponding collinear correlation function:
\begin{align}
\tilde{h}_{1L}^{\perp (\beta) q/N}(z,b;\mu)
&=ib^\beta M_N \tilde{h}_{1L}^{\perp(1)}(x,\mu),\label{eq:h_1Lb}\\
\tilde{H}_{1}^{\perp (\alpha) \pi/q}(z,b;\mu)
&=\frac{ib^\alpha}{2}\hat{H}^{(3)}(z,z,\mu),\label{eq:H_1b}
\end{align}
where the hard coefficients are calculated up to LO, and the longi-transversity of the nucleon target and the Collins FF of the pion in the $b$-space are defined as
\begin{align}
\tilde{h}_{1L}^{\perp \beta,{q/N}}(x,b;\mu)=&\int{d^2\bm{p}_T e^{-i\bm{p}_T\cdot b} \frac{\bm{p}_T^\beta}{M_N} h_{1L}^{\perp {q/N}}(x,\bm{p}_T^2;\mu)},\nonumber\\
\tilde{H}_{1}^{\perp \alpha, \pi/q}(z,b;\mu)=&\int{d^2\bm{K}_\perp e^{-i\bm{K}_\perp\cdot b/z} \frac{\bm{K}_\perp^\alpha}{M_\pi} H_{1}^{\perp \pi/q}(z,\bm{K}_\perp^2;\mu)}
\label{eq:polarizing FF}
\end{align}

The collinear function $\hat{H}^{(3)}(z,z,\mu)$ are the twist-3 quark-gluon-quark correlation function, which is related to the first $k_T$-moment of the Collins FF $H_{1}^{\perp(1)}$ by ~\cite{Yuan:2009dw}:
\begin{align}
\hat{H}^{(3)}(z,z,\mu)=\int{d^2\bm{K}_\perp\frac{|\bm{K}_\perp^2|}{M_\Lambda}H_{1}^{\perp {\pi}/q}(z,\bm{K}_\perp^2,\mu)}=2M_{\pi}\tilde{H}_{1}^{\perp(1)}(z,\mu).
\label{twist-3 Collins FF}
\end{align}

As for the nonperturbative part of the Sudakov form factor associated with the longi-transversity  and the Collins function, the information still remains unknown.
In a practical calculation, we assume that they are respectively the same as $S^\mathrm{pdf}_{\mathrm{NP}}$ and $S^\mathrm{ff}_{\mathrm{NP}}$.
Therefore, we can obtain the the longi-transversity and the Collins function in $b$-space as
\begin{align}
\tilde{h}_{1L}^{\perp (\beta) q/N}(x,b;Q)
&=ib^\beta M_N e^{-\frac{1}{2}S_{P}(Q,b_\ast)-S^\mathrm{pdf}_{\mathrm{NP}}(Q,b)}
\tilde{h}_{1L}^{\perp(1)}(x,\mu),\label{eq:h_1Lb}\\
\tilde{H}_{1}^{\perp (\alpha) \pi/q}(z,b;Q)
&=\frac{ib^\alpha}{2}e^{-\frac{1}{2}S_{\mathrm{P}}(Q,b_\ast)-S^{ff}_{\mathrm{NP}}(Q,b)}
\hat{H}^{(3)}(z,z,\mu).\label{eq:H_1b}
\end{align}

After performing the Fourier transformation, one can obtain the TMDs in the transverse momentum space
\begin{align}
\frac{p_{\perp}^\beta}{M_N}\tilde{h}_{1L}^{\perp q/N(\beta)}(x,p_T;Q)&=M_N \int_0^\infty\frac{db b^2}{2\pi}J_1(|\bm p_T|  b)e^{-\frac{1}{2}S_{P}(Q,b_\ast)-S^\mathrm{pdf}_{\mathrm{NP}}(Q,b)} \tilde{h}_{1L}^{\perp(1)}(x,\mu),\\
\frac{K_\perp^\alpha}{M_\pi}\tilde{H}_{1}^{\perp \pi/q(\alpha)}(z,K_\perp;Q)&=\int_0^\infty\frac{db b^2}{4\pi}J_1(|\bm K_\perp|b/z)e^{-\frac{1}{2}S_{P}(Q,b_\ast)-S^\mathrm{ff}_{\mathrm{NP}}(Q,b)} \hat{H}^{(3)}(z,z,\mu).
\end{align}

Besides the EIKV parametrization on non-perturbative factor $S_\mathrm{NP}$ mentioned above, another parametrization applied in this study is the Bacchetta, Delcarro, Pisano, Radici and Signori (BDPRS) parameterization $S_\mathrm{NP}$ for the unpolarized TMDs, which has the following form~\cite{Bacchetta:2017gcc}:
\begin{align}
\tilde{f}^{a}_1(x,b;Q)&=f^{a}_1(x;\mu^2)e^{-S(\mu,Q)}
e^{\frac{1}{2}g_K(b)\textrm{ln}(Q^2/Q_0^2)}\tilde{f}^{a}_{1\mathrm{NP}}(x,b),\label{eq:SNP-jhep-f}\\
\tilde{D}^{a\rightarrow h}_1(z,b;Q)&=D^{a\rightarrow h}_1(z;\mu^2)e^{-S(\mu,Q)}e^{\frac{1}{2}g_K(b)\textrm{ln}(Q^2/Q_0^2)}\tilde{D}^{a\rightarrow h}_{1\mathrm{NP}}(z,b),\label{eq:SNP-jhep1}
\end{align}
where $g_K=-g_2b^2/2$, following the choice in Refs.~\cite{Landry:2002ix,Nadolsky:1999kb,Konychev:2005iy}. $\tilde{f}^{a}_{1\mathrm{NP}}(x,b^2)$ and $\tilde{D}^{a\rightarrow h}_{1\mathrm{NP}}(z,b^2)$ are the intrinsic nonperturbative part of the PDFs and FFs respectively, which are parameterized as
\begin{equation}
\label{eq:SNP-jhep2-f}
\tilde{f}^a_{1\mathrm{NP}}(x,b^2)=\frac{1}{2\pi}e^{-g_{1}\frac{b^2}{4}}\left(1-\frac{\lambda g^2_{1}}{1+\lambda g_{1}}\frac{b^2}{4}\right),
\end{equation}
\begin{equation}
\label{eq:SNP-jhep2}
\tilde{D}^{a\rightarrow h}_{1\mathrm{NP}}(z,b^2)=\frac{g_3e^{-g_3\frac{b^2}{4z^2}}+(\frac{\lambda_F}{z^2})g_4^2(1-g_4\frac{b^2}{4z^2})e^{-g_4\frac{b^2}{4z^2}}}
{2\pi z^2 (g_3+(\frac{\lambda_F}{z^2})g_4^2)},
\end{equation}
with
\begin{equation}
g_{1}(x)=N_{1}\frac{(1-x)^\alpha x^\sigma}{(1-\hat{x})^\alpha \hat{x}^\sigma},
\end{equation}
\begin{equation}
g_{3,4}(z)=N_{3,4}\frac{(z^\beta+\delta)(1-z)^\gamma}{(\hat{z}^\beta+\delta)(1-\hat{z})^\gamma}.
\end{equation}
Here, $\hat{x}=0.1$ and $\hat{z}=0.5$ are fixed, and $\alpha, \sigma, \beta, \gamma, \delta$, $N_1\equiv g_1(\hat{x})$, $N_{3,4}\equiv g_{3,4}(\hat{z})$ are free parameters fitted to the available data from SIDIS, Drell-Yan, and $W/Z$ boson production processes.
Besides the $b_\ast(b)$ prescription in the original CSS approach~\cite{Collins:1984kg}, there are also several different choices on the form of $b_\ast(b)$~\cite{Collins:2016hqq,Bacchetta:2017gcc}.
In Ref.~\cite{Bacchetta:2017gcc}, a new $b_\ast$ prescription different from Eq.~(\ref{eq:b*}) was proposed as
\begin{equation}
b_\ast=b_\mathrm{max}\left(\frac{1-e^{{-b^4}/{b_\mathrm{max}^4}}}{1-e^{{-b^4}/{b_\mathrm{min}^4}}}\right)^{1/4}
\label{b*2}
\end{equation}
Again, $b_\mathrm{max}$ is the boundary of the nonperturbative and perturbative $b$ space region with fixed value of $b_\mathrm{max}=2e^{-\gamma _E}\ \textrm{GeV}^{-1}\approx 1.123\ \textrm{GeV}^{-1}$.
Furthermore, the authors in Ref.~\cite{Bacchetta:2017gcc} also chose to saturate $b_\ast$ at the minimum value $b_\mathrm{min}\propto 2e^{-\gamma_E}/Q$.

\section{The formalism of $\sin2\phi_h$ asymmetry of pion production in SIDIS}
\label{Sec.formalism}
In this section we will set up the necessary framework for physical observables in SIDIS process within TMD factorization by considering the evolution effects of TMDs.
The process under study is:
\begin{align}
l(\ell)+N^\rightarrow(P)\rightarrow l(\ell^\prime)+\pi(P_\pi)+X,
\end{align}
the lepton beam with momentum $\ell$ scatters off a longitudinally polarized nucleon target $N$ with momentum $P$.
In the final state, the scattered lepton momentum $\ell^\prime$ is measured together with an unpolarized final state hadron $h$ (in this work $h$ is the $\pi$ meson).
We define the space-like momentum transfer $q=\ell-\ell^\prime$ and introduce the relevant kinematic invariants
\begin{align}
x=\frac{Q^2}{2P\cdot q},~~~y=\frac{P\cdot q}{P\cdot \ell}=\frac{Q^2}{x_Bs},~~~z=\frac{P\cdot P_\pi}{P\cdot q},~~~Q^2=-q^2,~~~s=(P+\ell)^2.
\end{align}
Here, $s$ is the total center of mass energy squared, $x$ is the Bjorken variable, $y$ is the inelasticity and $z$ is the momentum fraction of the final state hadron.

\begin{figure}
  \includegraphics[width=0.5\columnwidth]{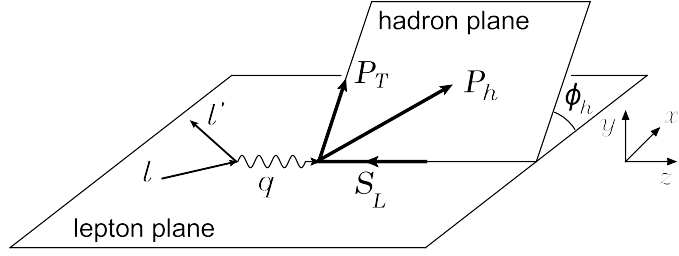}
 \caption{The kinematical configuration for the SIDIS process. The initial and scattered leptonic momenta define the lepton plane ($x-z$ plane), while the detected hadron momentum together with the $z$ axis identify the hadron production plane, the longitudinal spin of the nucleon is along the $-z$ axis.}
 \label{fig:SIDISframe}
\end{figure}

To leading order in $1/Q$ the SIDIS cross-section~(for a longitudinally polarized nucleon target) is given by~\cite{Bacchetta:2006tn}
\begin{align}
\frac{d^{5}\sigma}{dx dy dz d\phi_{\pi} dP_{\pi T}^2}
=\frac{2\pi\alpha^2}{x_ByQ^2}\times \left\{(1-y+\frac{1}{2}y^2)F_{UU}+(1-y) S_L \sin2\phi_{\pi} F_{UL}^{\sin2\phi_{\pi}}]\right\}.
\label{cross-section}
\end{align}
The corresponding reference frame of the process is given in Fig.~\ref{fig:SIDISframe} is applied.
In this frame, the virtual photon momentum $q$ defines the $z$-axis, the hadron plane is determined by the $z$-axis and the momentum direction of the final-state hadron, and the lepton plane is given by $\ell$ and $\ell^\prime$.
Hence $\phi_h$ is the azimuthal angle of the final-state hadron with respect to the lepton plane, respectively; $\bm{P}_{\pi T}$ is the component of $\bm{P}_\pi$ transverse with respect to $\bm q$ with $\bm{P}_{\pi T}=-z\bm{q}_T$~\cite{Boer:2011fh}.
Here, $P_{\pi T}$ is the characteristic transverse momentum detected in the SIDIS process, the value of which determines the validity of TMD factorization, i.e., if $\bm{P}_{\pi T}^2/z^2\ll Q^2$, TMD factorization can be applied and the process is sensitive to the TMDs~\cite{Collins:2011zzd}.
$F_{UU}$ and $F_{UL}^{\sin2\phi_h}$ are the spin-averaged and spin-dependent structure functions.

The expression for the $\sin2\phi_{h}$ azimuthal asymmetry is given by
\begin{align}
A_{UL}^{\sin 2\phi_h}(x,y,z,P_{\pi T})=\frac{\frac{1}{xyQ^2}(1-y)F_{UL}^{\sin2\phi_{h}}}
{\frac{1}{xyQ^2}(1-y+\frac{1}{2}y^2)F_{UU}}.
\label{A_UL}
\end{align}
According to TMD factorization, the structure functions $F_{UU}$ and  $F_{UL}^{\sin2\phi_{\pi}}$ are given in terms of an integral which convolutes transverse parton momentum in the distribution and the fragmentation function~\cite{Kotzinian:1994dv,Kotzinian:1997wt}
\begin{align}
F_{UU}(x,Q;P_{\pi T}^2)
=&x\sum_q e_q^2 \int d^2\bm{p}_T d^2\bm{k}_T \delta^2(\bm{p}_T-\bm{k}_T+\bm{q}_{T})f_1^{q/N}(x,\bm{p}_T^2;Q)D_1^{\pi/q}(z,\bm{k}_T^2;Q),
\label{F_UU}\\
F_{UL}^{\sin2\phi_h}(x,Q;P_{\pi T})
=&x\sum_q e_q^2 \int d^2\bm{p}_T d^2\bm{k}_T \delta^2(\bm{p}_T-\bm{k}_T+\bm{q}_T)[-\frac{2(\bm{\hat{h}}\cdot\bm{k}_{T})(\bm{\hat{h}}\cdot\bm{p}_{T})-\bm{k}_{T}\cdot\bm{p}_{T}}
{M_N M_\pi}]\nonumber\\
&\times h_{1L}^{\perp {q/N}}(x,\bm{p}_T^2;Q) H_{1}^{\perp \pi/q}(z,\bm{k}_T^2;Q)
\label{F_UL}
\end{align}
where the unit vector $\bm{\hat{h}}$ is defined as $\bm{\hat{h}}=\bm{P}_{\pi T}/P_{\pi T}$,
and the transverse momentum $\bm{k}_T$ is related to the transverse momentum of the produced hadron with respect to the quark through $\bm{K}_\perp=-z\bm{k}_T$.

It is convenient to deal with the TMD evolution effect in the $b$ space that is conjugate to the transverse momentum space through Fourier transformation, since it can turn the complicated convolution in the transverse momentum space into simple product.
Therefore, we perform a transformation for the delta function
\begin{align}
\delta^{2}(\bm{p}_T-\bm{k}_T+\bm{q}_T) = {1\over (2\pi)^2}\int d^2 \bm{b}_\perp e^{-i \bm b_\perp\cdot(
\bm{p}_T-\bm{k}_T+\bm{q}_T)},
\end{align}
and obtain the following explicit form of the spin-averaged structure function $F_{UU}$
\begin{align}
&F_{UU}(x,Q;P_{\pi T})\nonumber\\
=&x\frac{1}{z^2}\sum_{q}e_q^2\int \frac{d^2b}{(2\pi)^2}
e^{i\bm{P}_{\pi T}\cdot b/z}\tilde{f}_1^{q/N}(x,b;Q)\tilde{D}_1^{\pi/q}(z,b;Q)\nonumber\\
=&x\frac{1}{z^2}\sum_{q}e_q^2\int \frac{dbb}{2\pi}J_0(|\bm{P}_{\pi T}||b|/z)
\tilde{f}_1^{q/N}(x,b;Q)\tilde{D}_1^{\pi/q}(z,b;Q).
\label{eq:FUU}
\end{align}
The unpolarized PDF and FF in $b$ space can be defined as~
\begin{align}
\tilde{f}_1^{q/N}(x,b;Q)&=\int{d^2\bm{p}_T e^{-i\bm{p}_T \cdot b} f_1^{q/N}(x,\bm{p}_T^2;Q)},\nonumber\\
\tilde{D}_1^{\pi/q}(z,b;Q)&=\int{d^2\bm{K}_\perp e^{-i\bm{K}_\perp \cdot b/z}
D_1^{\pi/q}(z,\bm{K}_\perp^2;Q)}.
\label{eq:Upolarizing FF}
\end{align}
Similarly, the spin-dependent structure function $F_{UL}^{\sin(2\phi_h)}$ can be written as
\begin{align}
&F_{UL}^{\sin2\phi_h}(x,Q;P_{\pi T})\nonumber\\
=&x\frac{1}{z^3}\sum_q e_q^2 \int d^2\bm{p}_T d^2\bm{K}_\perp
\int \frac{d^2b}{(2\pi)^2} e^{-i(\bm{p}_T+{\bm{K}_\perp}/z-{\bm{P}_{\pi T}}/z)\cdot b}
[2(\bm{\hat{h}}\cdot\bm{K}_{\perp})(\bm{\hat{h}}\cdot\bm{p}_{T})-\bm{K}_{\perp}\cdot\bm{p}_{T}]\nonumber\\
&\times\frac{h_{1L}^{\perp {q/N}}(x,\bm{p}_T^2;Q) H_{1}^{\perp \pi/q}(z,\bm{K}_\perp^2;Q)}
{M_N M_\pi}\nonumber\\
=&x\frac{1}{z^3}\sum_q e_q^2
\int \frac{d^2b}{(2\pi)^2} e^{i{\bm{P}_{\pi T}}\cdot b /z}
(2\hat{h}_\alpha\hat{h}_\beta-g_{\alpha\beta})\tilde{h}_{1L}^{\perp \beta, {q/N}}(x,b;Q) \tilde{H}_{1}^{\perp \alpha, \pi/q}(z,b;Q)\nonumber\\
=&x\frac{1}{z^3}\sum_q e_q^2\int \frac{dbb^3}{4\pi}J_2(|\bm{P}_{\pi T}||b|/z)M_N
\tilde{h}_{1L}^{\perp (1), {q/N}}(x;\mu) \hat{H}^{ (3), \pi/q}(z,z;\mu)e^{-(S _{P}(Q,b_\ast)+S^\mathrm{pdf}_{\mathrm{NP}}(Q,b)+S^{ff}_{\mathrm{NP}}(Q,b))}.
\label{eq:FUT}
\end{align}
Therefore, we obtain the evolved form of the structure function $F_{UL}^{\sin2\phi_h}(x,Q;P_{\pi T})$.

\section{Numerical calculation}

\label{Sec.numerical}
\begin{figure}
  \centering
  \includegraphics[width=0.4\columnwidth]{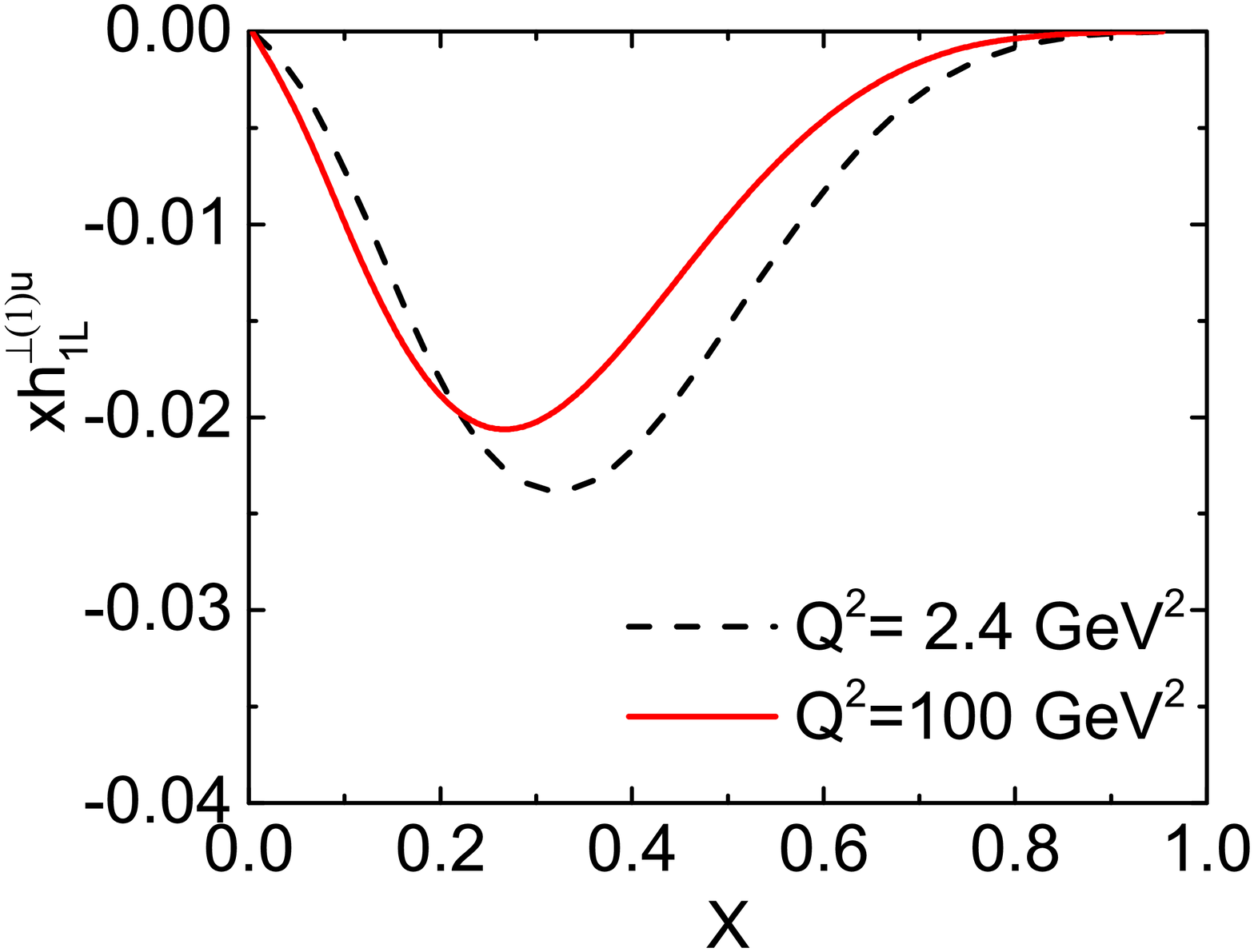}
  \includegraphics[width=0.4\columnwidth]{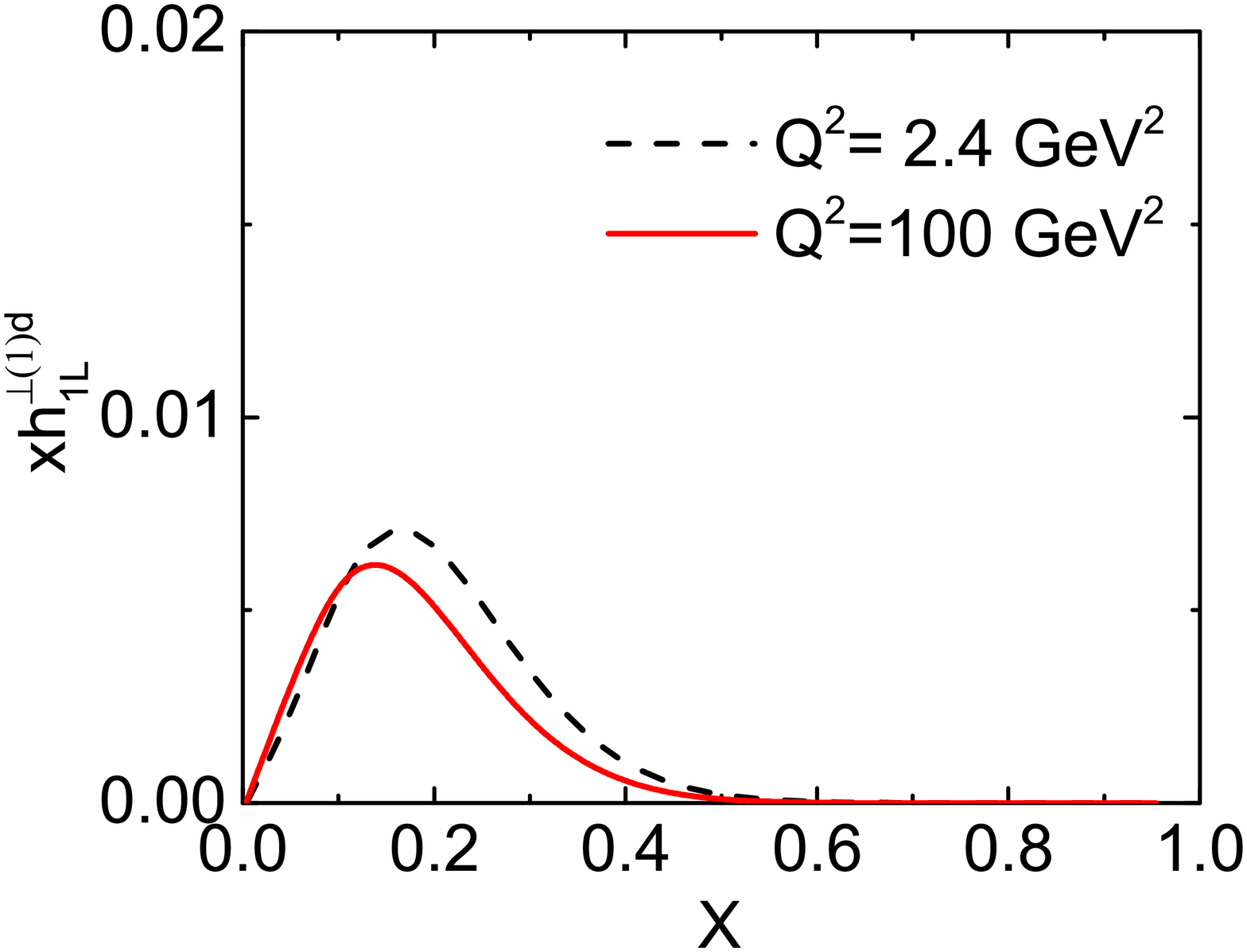}
  \caption{Left panel: $h_{1L}^{\perp (1)}(x,Q^2)$ (multiplied by x) of the proton longitudinal transversity PDF $h_{1L}^{\perp}$ for up quark at $Q^2=2.41\ \textrm{GeV}^2$ and $Q^2=10\ \textrm{GeV}^2$. Right panel: similar to the left panel, but for the down quark.}
  \label{fig1:xh1l}
\end{figure}
Using the framework set up above, in this section, we present the numerical calculation of the $\sin2\phi_h$ azimuthal asymmetry for $\pi$ production in the process $lN^\rightarrow\to l\pi X$ off the longitudinally polarized nucleon.
We estimate the asymmetry at the kinematical configurations of HERMES, CLAS and CLAS12, and compare it with the recent experimental measurements~\cite{Airapetian:1999tv, Airapetian:2001eg,Airapetian:2002mf,Jawalkar:2017ube,Avakian:2010ae,Diehl:2021rnj,Parsamyan:2018ovx, Parsamyan:2018evv,Adolph:2016vou,Alekseev:2010dm}.
For this purpose we need to know the collinear functions appearing in Eqs.~(\ref{eq:f_b}), (\ref{eq:D_b}), (\ref{eq:h_1Lb}) and (\ref{eq:H_1b}).
For the unpolarized PDF $f_1(x, \mu)$ of the nucleon, we apply the NLO set of the CT10 parametrization (central PDF set)~\cite{Lai:2010vv}. For the unpolarized FF $D_1(z, \mu)$ of the pion, we apply the NLO set of the de Florian, Sassot, Stratmann (DSS) FF~\cite{deFlorian:2007aj}.
For the twist-3 Collins FF $H^{(3)}(z,z,\mu)$ of the $\pi$, we adopt the parametrization from Ref.~\cite{Kang:2015msa}. For the first transverse moment of the longitudinal transversity $\tilde{h}_{1L}^{\perp(1)}(x,\mu)$ of the nucleon, we adopt the WW-type approximations from Ref.~\cite{Bastami:2018xqd}.

In the WW-approximation one can write $h_{1L}^{\perp (1)}(x)$ as a total derivative expressed in terms of the transversity distribution $h_1(x)$ as follows~\cite{Bastami:2018xqd}
\begin{align}
h_{1L}^{\perp (1)}(x)\overset{WW-type}{\approx} -x^2\int_{x}^{1}\frac{dy}{y^2}h_1^a(y).
\label{WW-type}
\end{align}
Here, we use the extractions of $h_1(x)$ from Ref.~\cite{Kang:2015msa}.
The WW-type approximation was discussed in Refs.~\cite{Mulders:1995dh,Kotzinian:1995cz,Kotzinian:1997wt,Kotzinian:2006dw,Avakian:2007mv,Metz:2008ib,Teckentrup:2009tk,Tangerman:1994bb} in details.

As for the energy evolution of twist-3 FF of quark flavor $q$ to $\pi$, $\hat{H}^{(3)}$, the evolution effect can be obtained from Ref.~\cite{Kang:2015msa}.
Moreover, the energy evolution for $\tilde{h}_{1L}^{\perp (1)}$ has been studied extensively in Ref.~\cite{Zhou:2008mz} and has a more complicated form.
In this paper, similar to the choice in Ref.~\cite{Kang:2015msa}, we only keep the homogeneous terms of the evolution kernel of $\tilde{h}_{1L}^{\perp (1)}$, which has the following form:
\begin{align}
P_{qq}^{\tilde{h}_{1L}^{\perp (1)}}=C_F\Big[\frac{2z}{(1-z)_+}+2\delta(1-z)\Big]-\frac{C_A}{2}\frac{2z}{1-z}.
\label{sf}
\end{align}

The numerical solution of DGLAP equations is performed by the QCDNUM evolution package~\cite{Botje:2010ay}.
The original code of QCDNUM is modified by us so that the evolution kernels of $\hat{H}_{1}^{\perp(3)}$ and $\tilde{h}_{1L}^{\perp (1)}$ are included.
In Fig.~\ref{fig1:xh1l}, we plot the $h_{1L}^{\perp (1)}(x,Q^2)$ (multiplied by $x$) vs $x$ for light quark flavors at the initial scale $Q^2$=2.4 GeV$^2$ as well as the evolved scale $Q^2 =100$ GeV$^2$.
The left panel and the right panel show the results for the up quark and the down quark, respectively.
The plots show that the $h_{1L}^{\perp(1)}(x,Q^2)$ for up quark is larger than one for the down quark in size, and with the opposite sign for them.
Also, the evolution effect from lower scale to higher scale drives the peak of the distribution to the smaller $x$ region.

The $A_{UL}^{\sin2\phi_h}$ asymmetry pion production was measured at HERMES off the proton target~\cite{Airapetian:1999tv, Airapetian:2001eg} and off the deuteron target~\cite{Airapetian:2002mf} in the kinematic range
\begin{align}
1\ \textrm{GeV}^2<Q^2<15\ \textrm{GeV}^2, \quad W>2\ \textrm{GeV},\quad 0.023<x<0.4, \quad 0.2<y<0.85,\quad 0.2<z<0.7,
\label{HERMES-kinematic range}
\end{align}
while it was also measured for pion production off the proton target by CLAS~\cite{Jawalkar:2017ube,Avakian:2010ae} in the kinematic range
\begin{align}
1\ \textrm{GeV}^2<Q^2<5.4\ \textrm{GeV}^2, \quad W>2\ \textrm{GeV},\quad 0.12<x<0.48, \quad y<0.85,\quad 0.4<z<0.7,
\label{HERMES-kinematic range}
\end{align}
with $W^2=(P+q)^2\approx\frac{1-x}{x}Q^2$ being the invariant mass of the virtual photon-nucleon system.

\begin{figure}
  \centering
  \includegraphics[width=0.4\columnwidth]{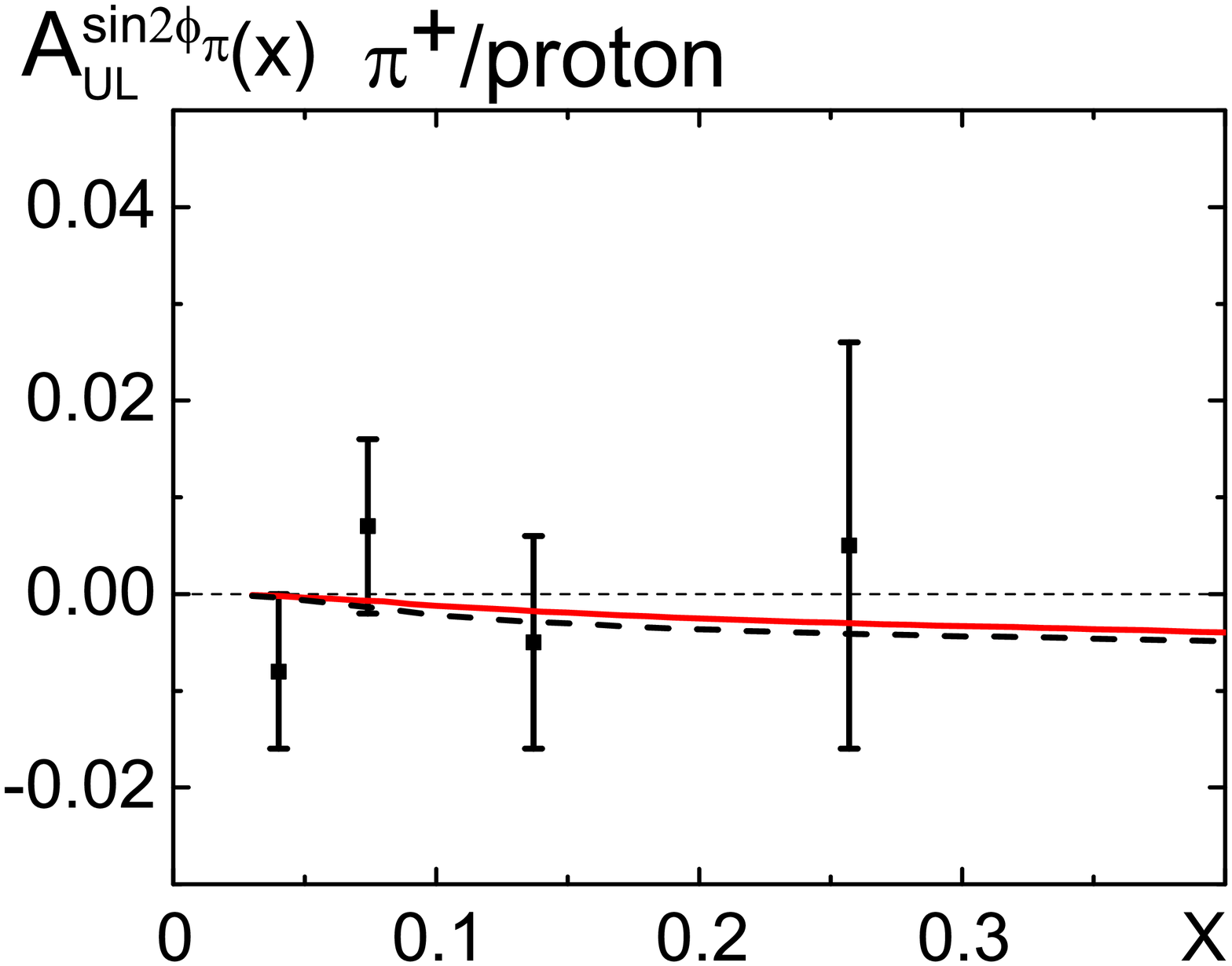}~~~~~~
     \includegraphics[width=0.4\columnwidth]{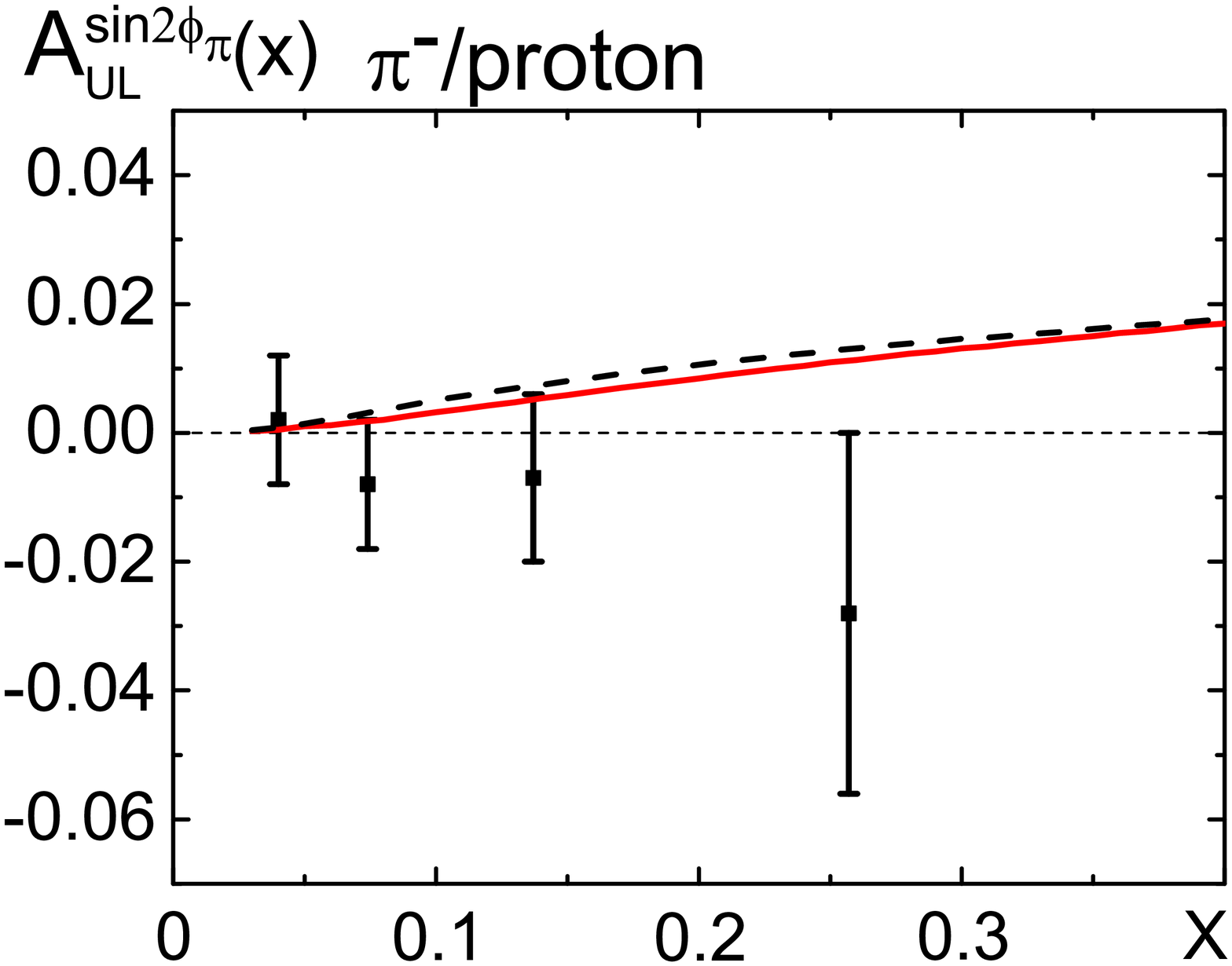}
  \caption{The  $A_{UL}^{\sin2\phi_h}$ asymmetry for the $\pi^+$ (left) and $\pi^-$ (right) productions off the proton target as function of $x$ at HERMES. The solid lines correspond to the results from the BDPRS parametrization~\cite{Bacchetta:2017gcc} [Eqs.~(\ref{eq:SNP-jhep-f}) and (\ref{eq:SNP-jhep1})] on the nonperturbative form factor, while the dashed lines correspond to the results calculated from the EIKV parametrization~\cite{Echevarria:2014xaa} [Eqs.~(\ref{eq:eikv1}) and (\ref{eq:eikv2})] on the nonperturbative form factor. The solid squares represent the HERMES data for comparison. In each figure we zoom in the plots to show the tendency better.}
  \label{fig1:HERMES}
\end{figure}
\begin{figure}
  \centering
  \includegraphics[width=0.32\columnwidth]{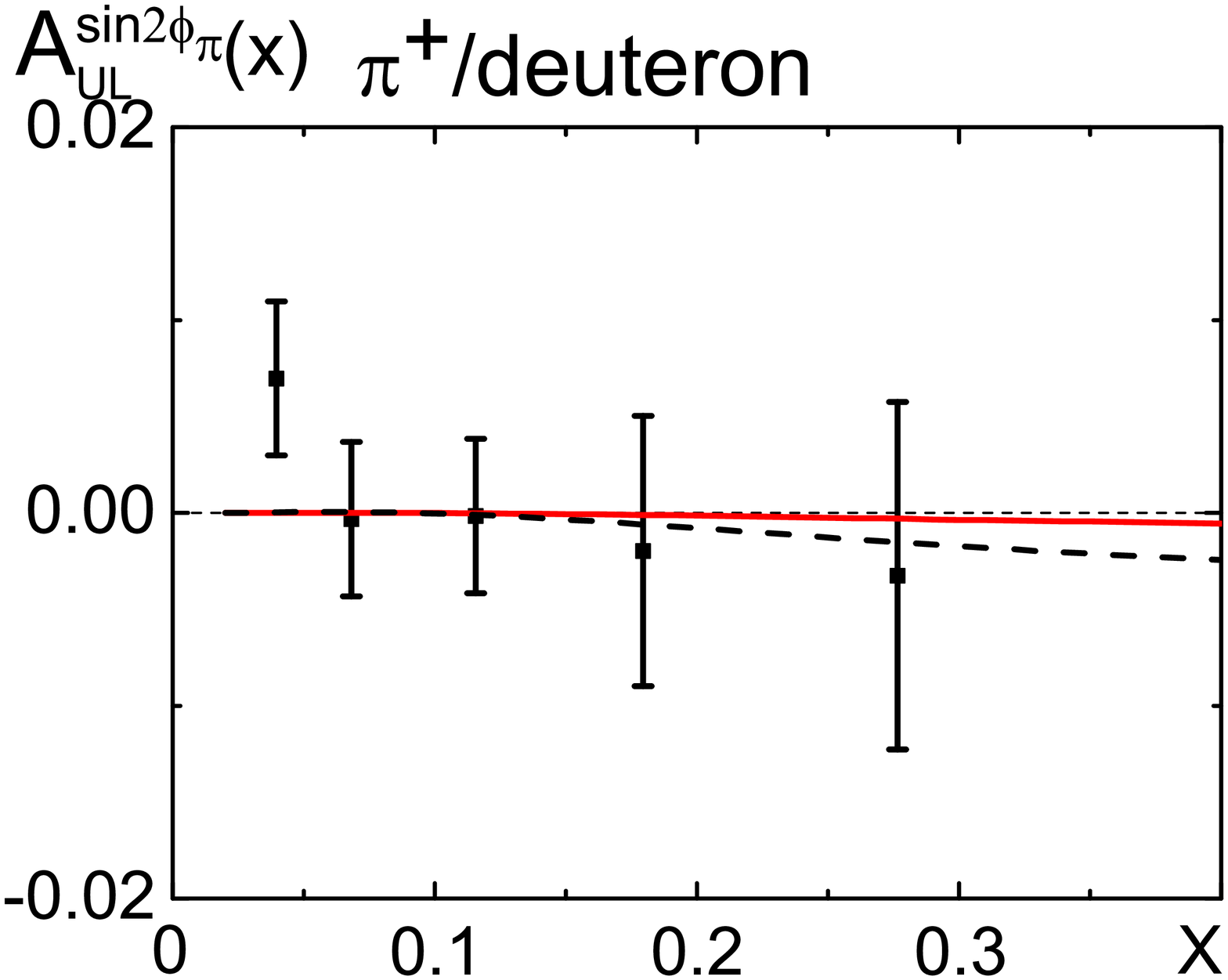}
  \includegraphics[width=0.32\columnwidth]{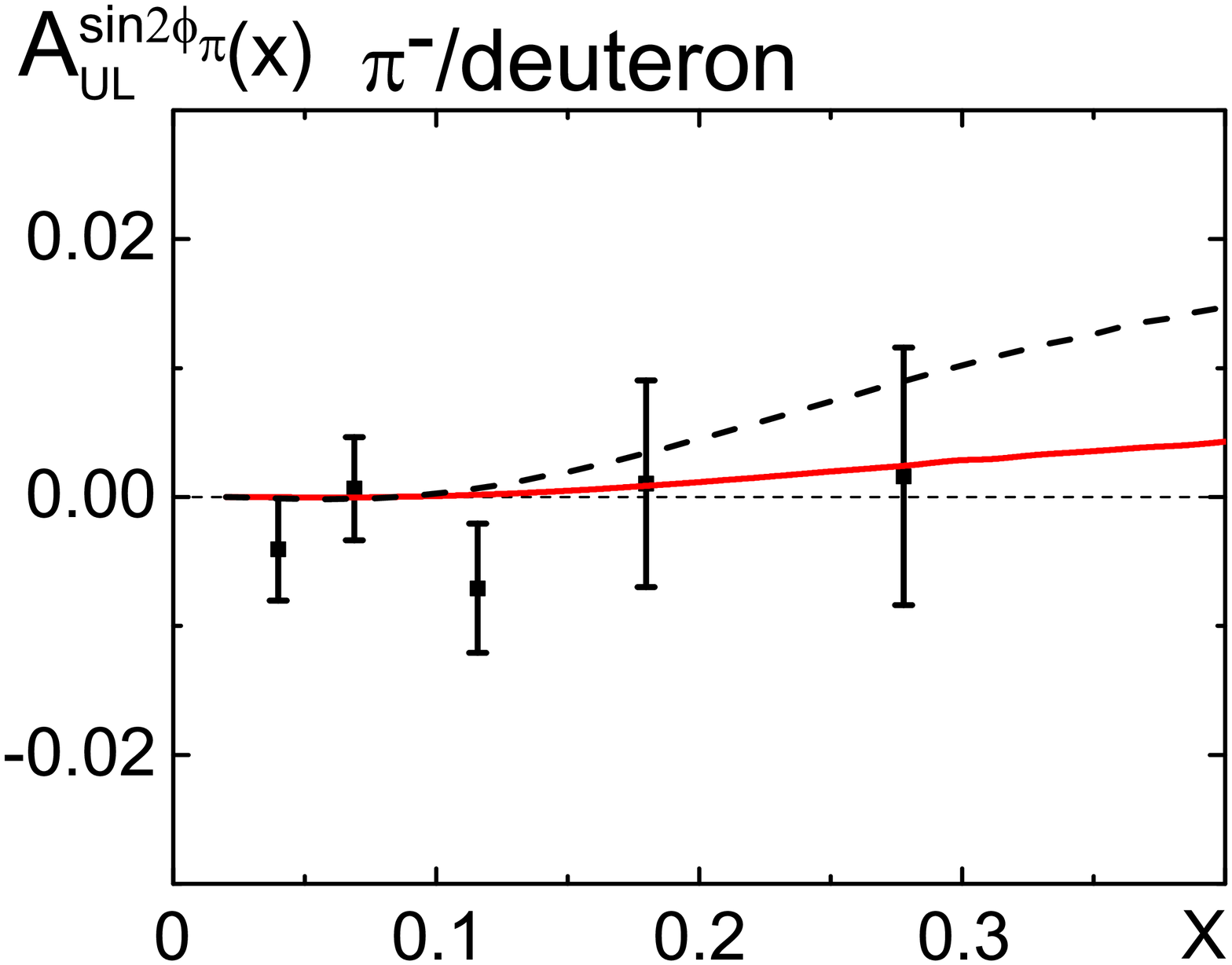}
  \includegraphics[width=0.32\columnwidth]{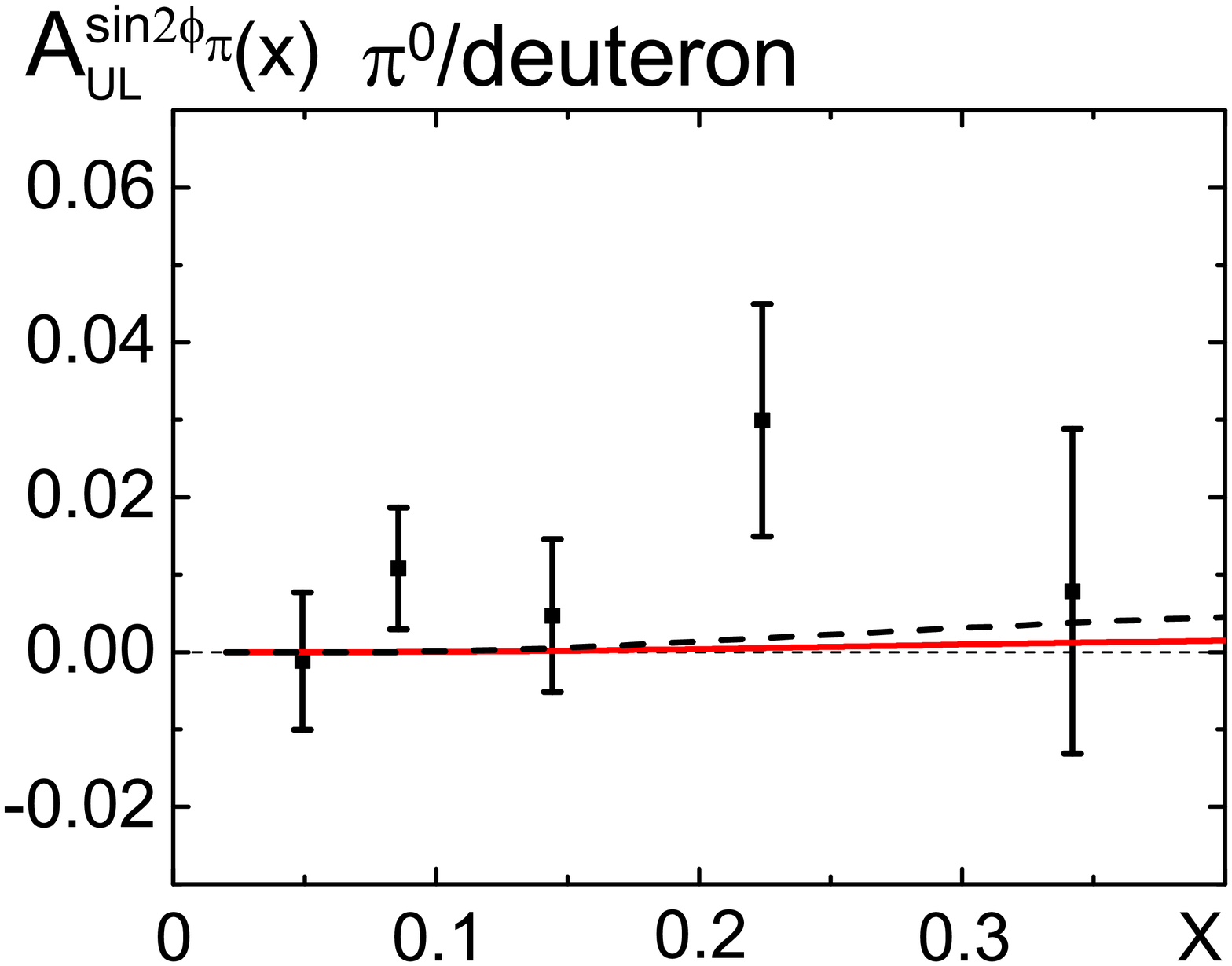}
  \caption{Similar to Fig.~\ref{fig1:HERMES}, but for $\pi^+$ (left), $\pi^-$ (central) and $\pi^0$ (right) productions off the deuteron target.}
  \label{fig2:HERMES2}
\end{figure}

In Figs.~\ref{fig1:HERMES} and \ref{fig2:HERMES2}, we plot our numerical results of the $\sin2\phi_h$ azimuthal asymmetry off the longitudinally polarized proton and deuteron target in the SIDIS process at HERMES kinematics~\cite{Avakian:1999rr,Avakian:2007mv,Airapetian:1999tv, Airapetian:2001eg,Airapetian:2002mf}, based on the TMD factorization formalism described in Eqs.~(\ref{A_UL}),~(\ref{eq:FUU}) and~(\ref{eq:FUT}). In this calculation we apply the EIKV parametrization (dark dashed line) and the BDPRS parametrization (red solid line) for the nonperturbative part with $b_\ast$ prescription in Eq.~(\ref{eq:b*}) and Eq.~(\ref{b*2}) respectively. To make the TMD factorization valid, the integration over the transverse momentum $P_{\pi T}$ is performed in the region of $P_{\pi T}<0.5\ \textrm{GeV}$.
In the figure the solid squares show the experimental data measured by the HERMES collaboration~\cite{Avakian:1999rr,Avakian:2007mv,Airapetian:1999tv, Airapetian:2001eg,Airapetian:2002mf}, with the error bars corresponding to the statistical uncertainty.

As shown in Figs.~\ref{fig1:HERMES} and \ref{fig2:HERMES2}, in all the cases the $\sin2\phi_h$ azimuthal asymmetries in the SIDIS process is rather small, around $2\% $ at most. The estimated $\sin2\phi_h$ azimuthal asymmetry for $\pi^+$ is negative, while those for $\pi^-$ and $\pi^0$ are positive. The size of the asymmetry for $\pi^-$ is larger than that for $\pi^+$ and $\pi^0$.
In addition, our estimates also show that the size of the asymmetries increase with increasing $x$. In the case of the proton target, the results form the EIKV parametrization are close to those from the BDPRS parametrization; while in the case of the deuteron target, there is quantitative difference between the two parametrization, namely, the results from the is two times larger than that from the BDPRS pamametrization. From the comparison with the HERMES data, we find that our results are consistent with the HERMES data~\cite{Avakian:1999rr,Avakian:2007mv,Airapetian:1999tv, Airapetian:2001eg,Airapetian:2002mf} within the error band, except the $\pi^-$ production at larger $x$ region, where the HERMES data show a negative asymmetry.
A nonzero $\sin2\phi_h$ azimuthal asymmetries for $\pi^{\pm}$ are estimated, indicating that spin-orbit correlation of transversely polarized quarks in the longitudinally polarized nucleon may be significant.

\begin{figure}
  \centering
  \includegraphics[width=0.4\columnwidth]{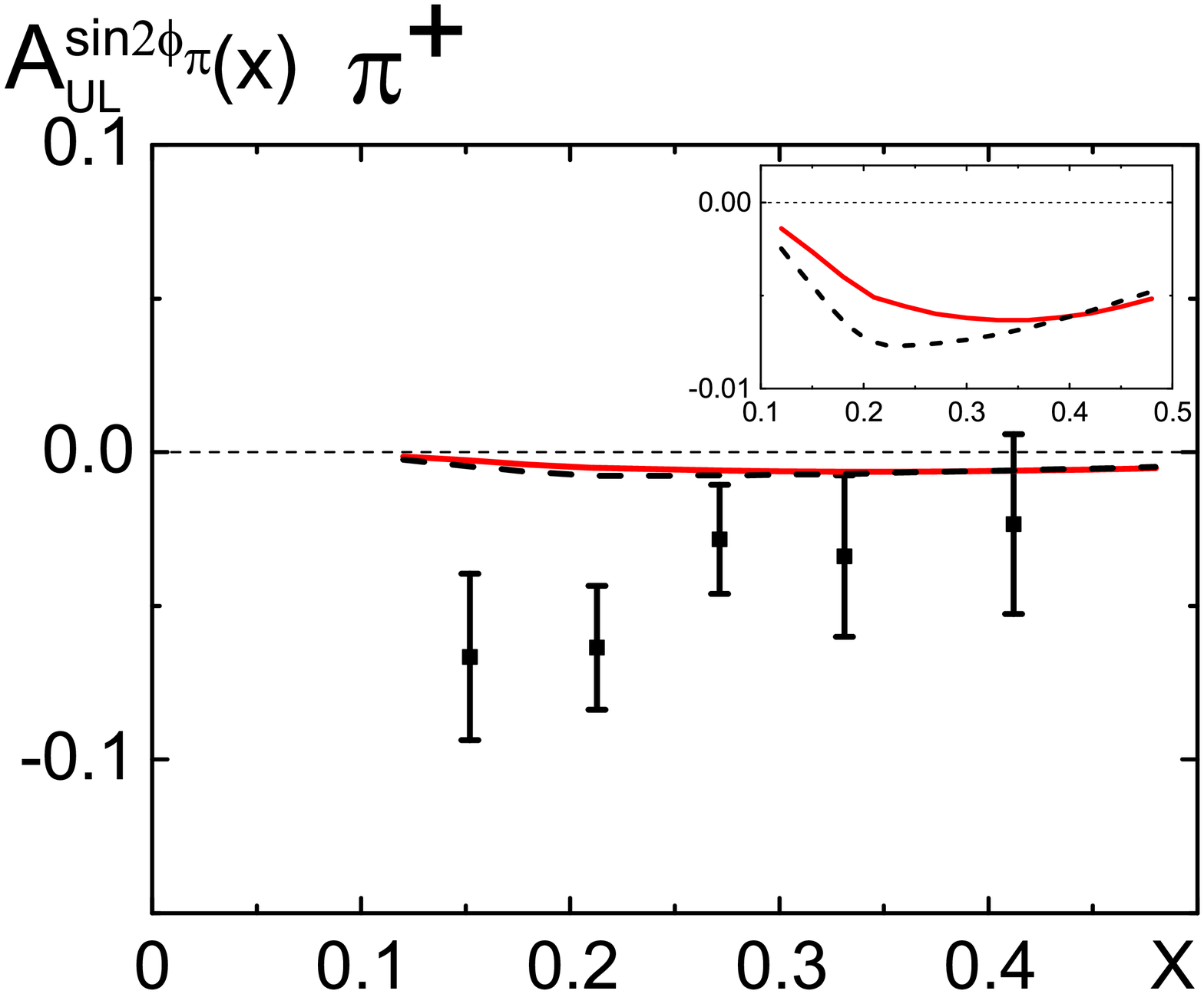}~~~~~
  \includegraphics[width=0.4\columnwidth]{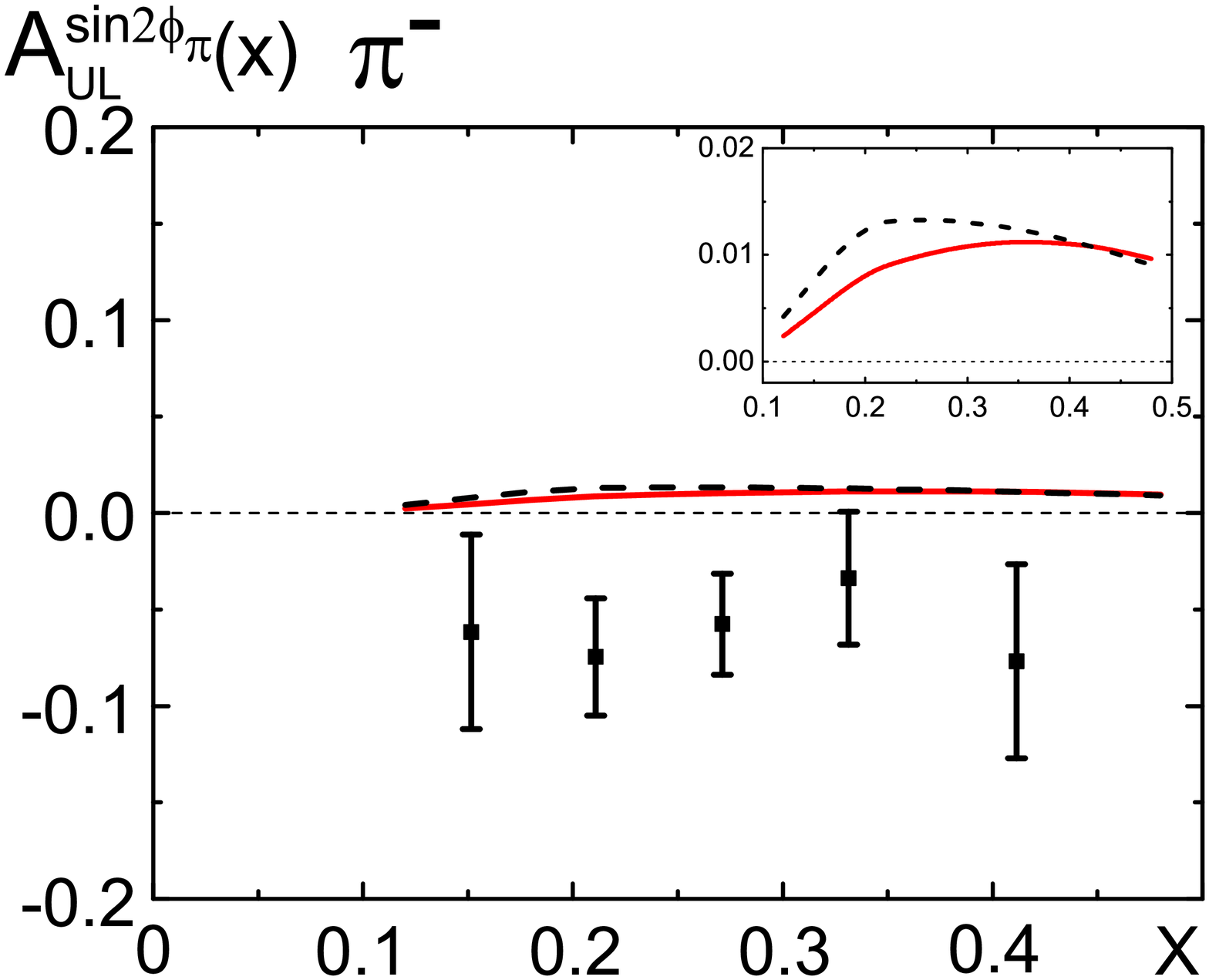}\\
  \vspace{0.4cm}
  \includegraphics[width=0.4\columnwidth]{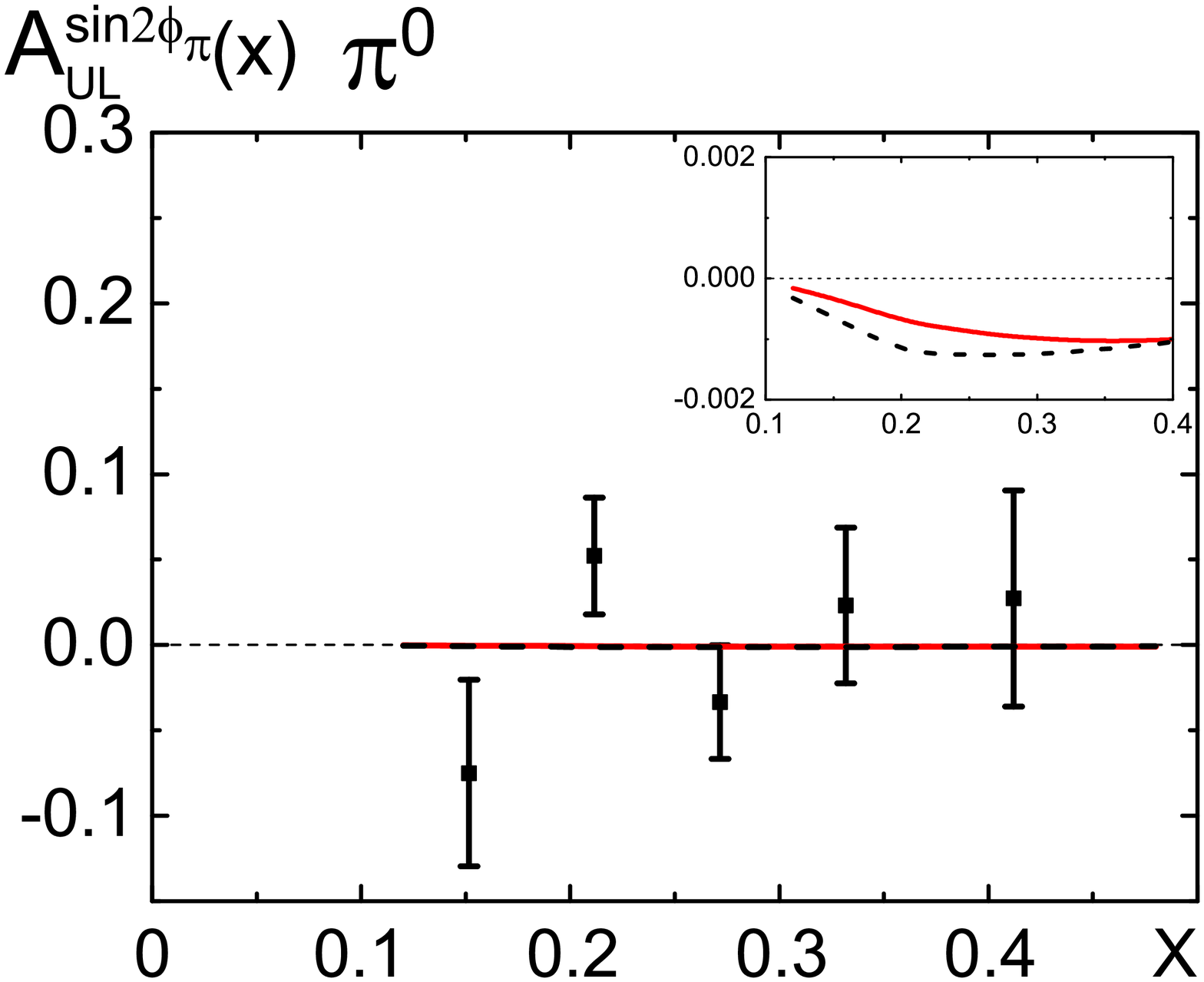}~~~~~
  \includegraphics[width=0.4\columnwidth]{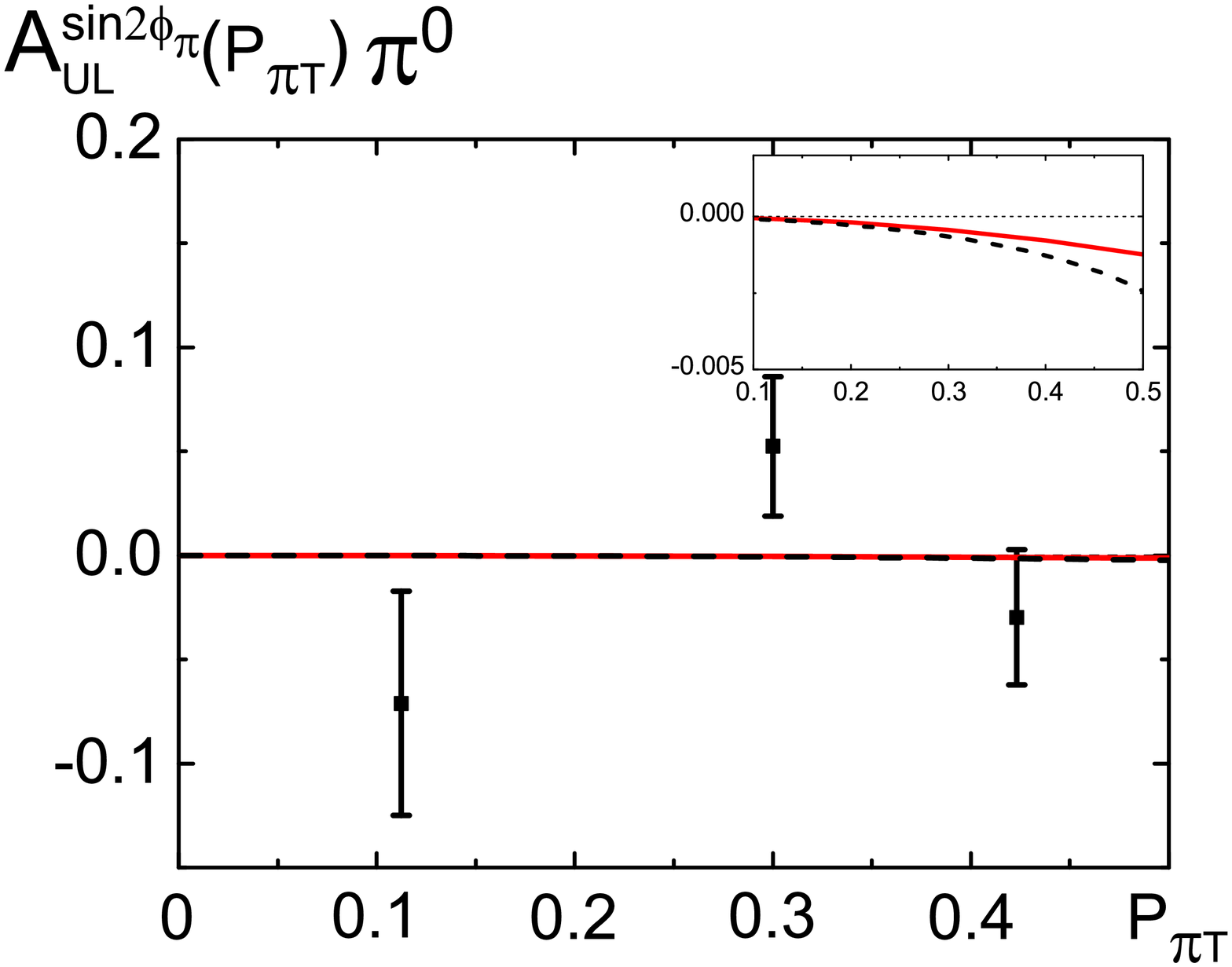}
  \caption{Similar to Fig.~\ref{fig1:HERMES}, but the Longitudinal proton target $A_{UL}^{\sin2\phi_h}$ as function of $x$ and $P_{\pi T}$ at CLAS kinematics. The upper panels of show the $x$ dependent asymmetries for $\pi^+$ (left panel) and $\pi^-$ (right panel) production  respectively; and the lower panels depict the $x-$ dependent (left panel) and $P_{\pi T}-$ dependent (right panel) asymmetries for $\pi^0$ production. The solid squares represent the CLAS data for comparison.}
  \label{fig2:CLAS}
\end{figure}

In Fig.~\ref{fig2:CLAS}, we plot our numerical results for the $\sin2\phi_h$ azimuthal asymmetry off the proton target, but at the CLAS kinematics~\cite{Jawalkar:2017ube,Avakian:2010ae}. The upper panels of Fig.~\ref{fig2:CLAS} show the $x$-dependent asymmetries of $\pi^+$ (left) and $\pi^-$ (right) production; and the lower panels depict the $x-$ dependent (left) and $P_{\pi T}-$ dependent (right) asymmetries of $\pi^0$ production.
In each figure we zoom in the plots to show the tendency better. 
Although the uncertainty at CLAS is relatively large, the data indicate that the asymmetries of both $\pi^+$ and $\pi^-$ at CLAS tend to be negative. Thus our estimate miss data of the x-dependent asymmetry for $\pi^-$ production.
A much smaller $\sin2\phi$ azimuthal asymmetries for $\pi^0$ production is predicted at CLAS as well as at HERMES off deuteron target, in agreement with experimental data. This is because the favored and unfavored Collins functions is summed for $\pi^0$ production, which means that they largely cancel for $\pi^0$.

Even though the $A_{UL}^{\sin2\phi_h}$ asymmetries off longitudinally polarised targets has been studied at HERMES (proton, deuteron)~\cite{Airapetian:1999tv, Airapetian:2001eg,Airapetian:2002mf}, CLAS (proton)~\cite{Jawalkar:2017ube,Avakian:2010ae} kinematics during the last two decades, there is still no consistent understanding of the contribution of each part to the total structure function which may be due to the low statistics or limited kinematic coverage of previous experiments.
Therefore, we also estimate the $A_{UL}^{\sin2\phi_h}$ asymmetries on proton target for pion production at the kinematical configuration of CLAS12 experiment~\cite{Diehl:2021rnj}
\begin{align}
1\ \textrm{GeV}^2<Q^2<7\ \textrm{GeV}^2, \quad W>2\ \textrm{GeV},\quad 0.13<x<0.52, \quad y<0.75,\quad 0.18<z<0.7
\label{HERMES-kinematic range}.
\end{align}
\begin{figure}
  \centering
  \includegraphics[width=0.329\columnwidth]{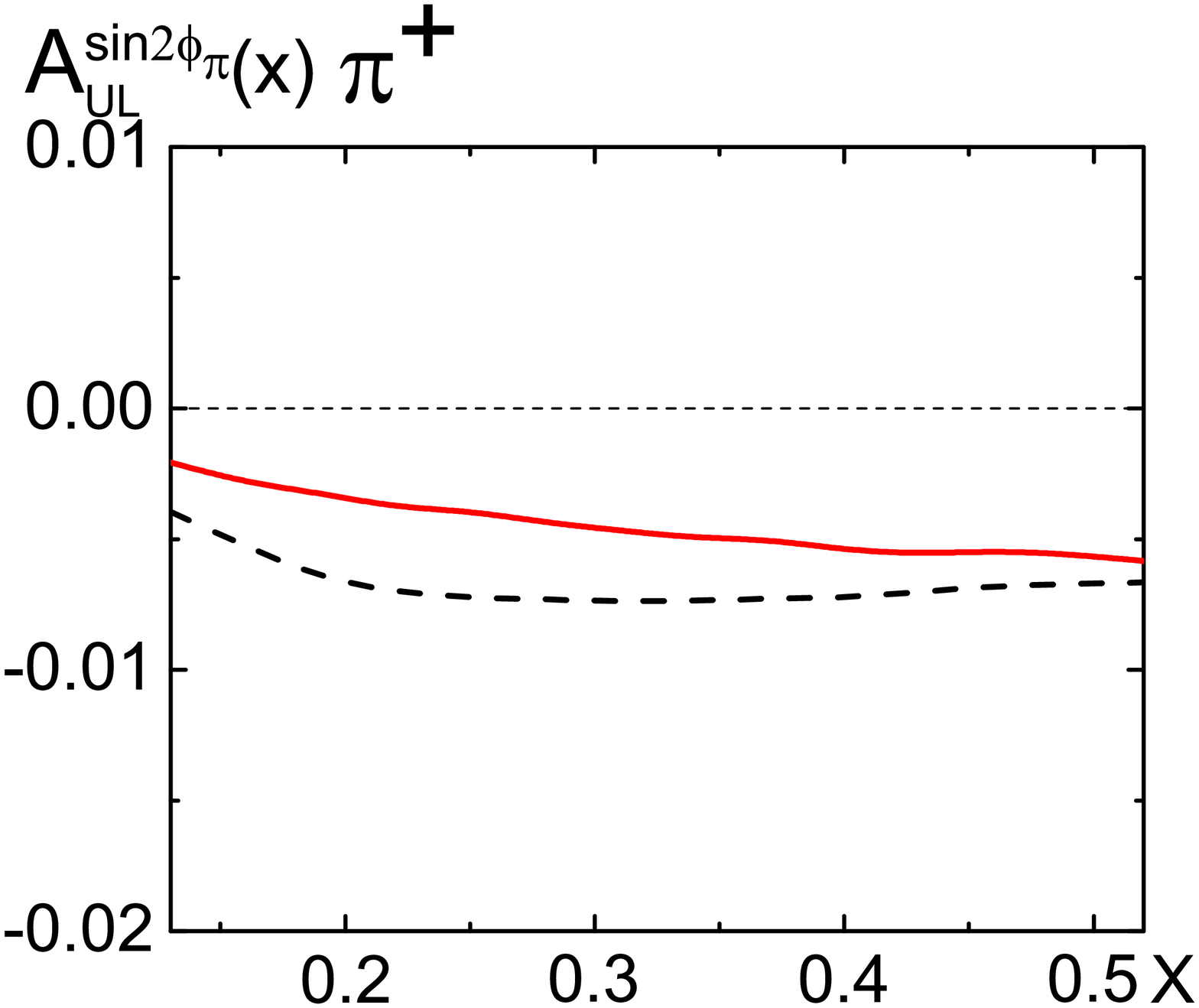}
  \includegraphics[width=0.329\columnwidth]{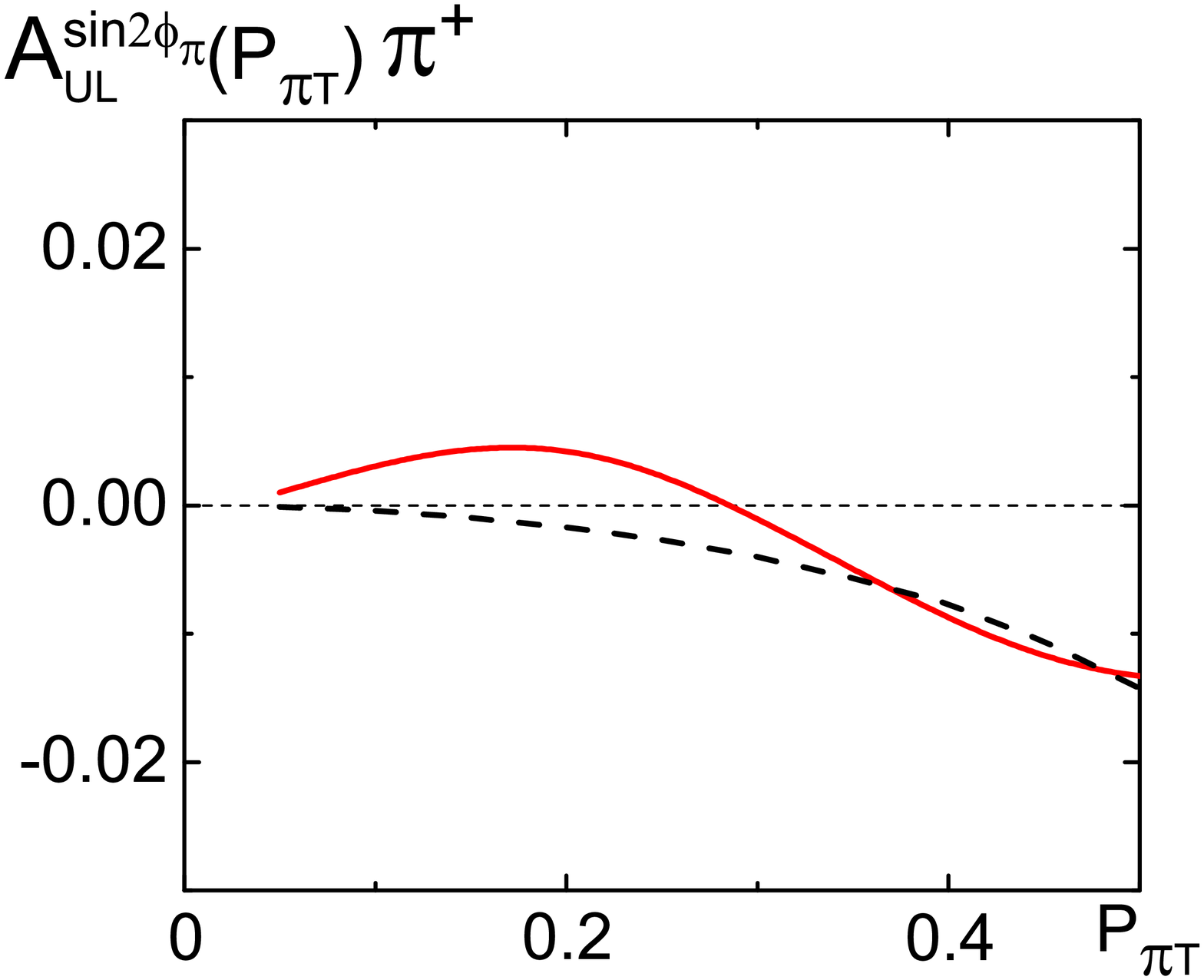}
  \includegraphics[width=0.329\columnwidth]{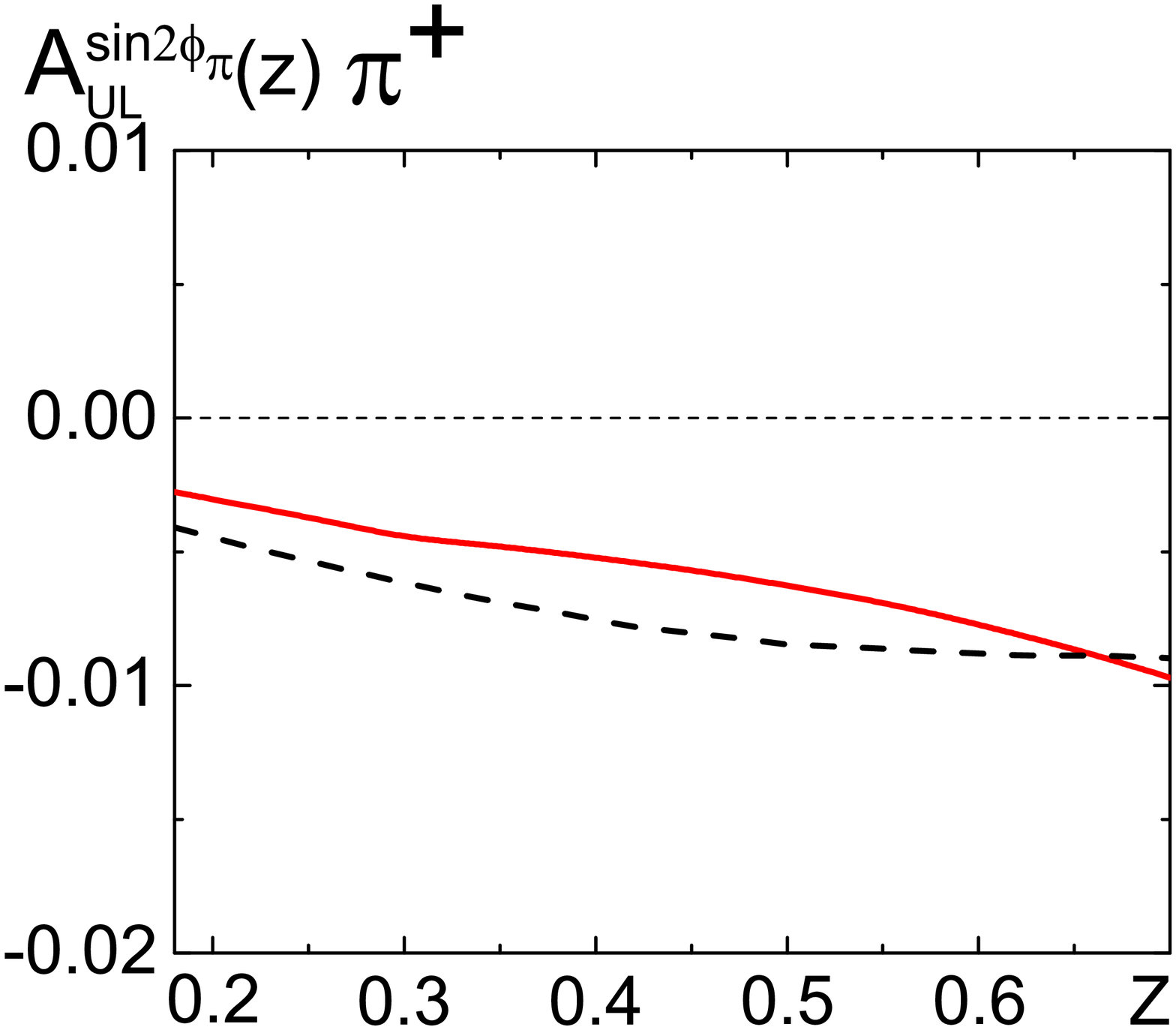}
  \includegraphics[width=0.329\columnwidth]{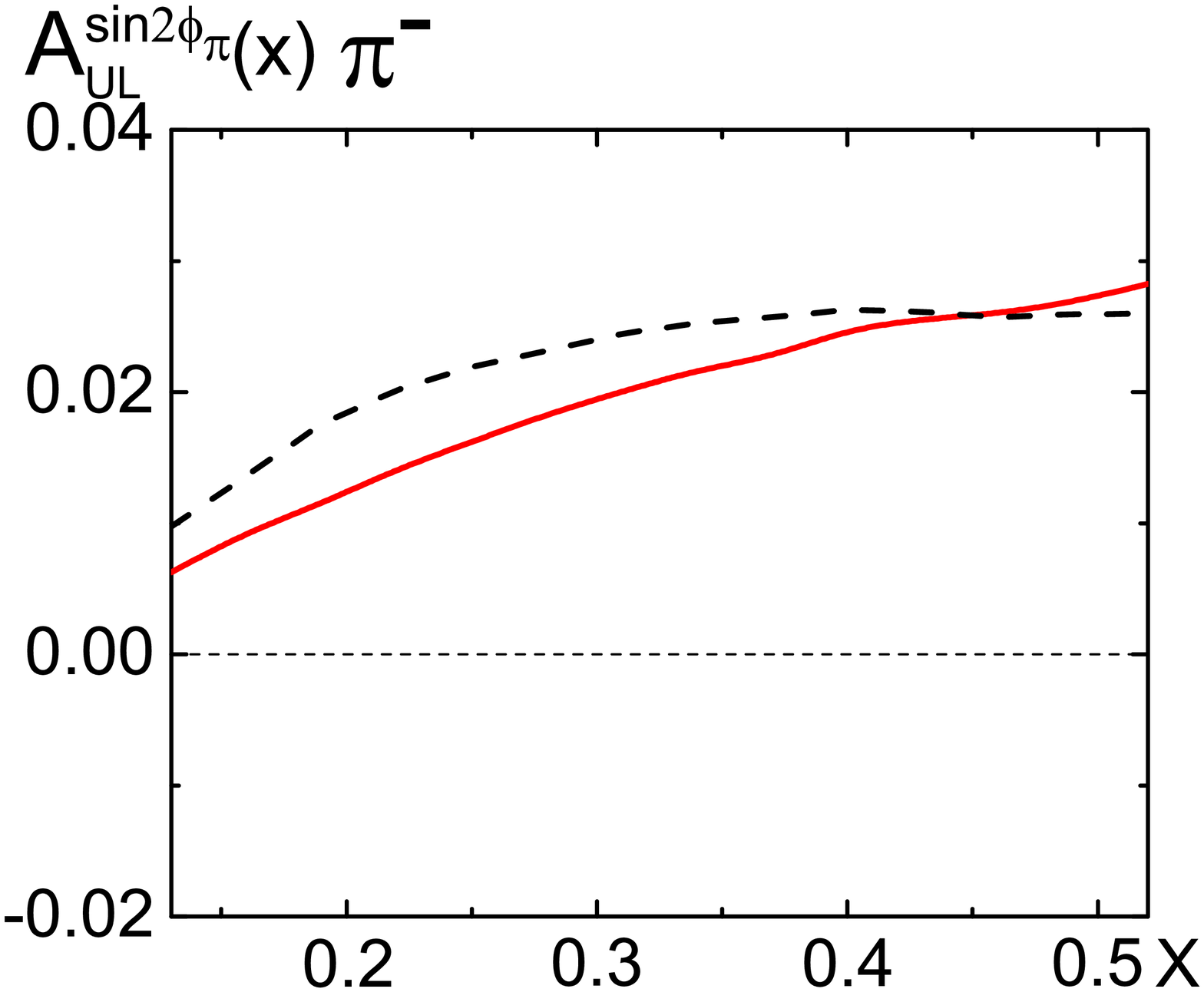}
  \includegraphics[width=0.329\columnwidth]{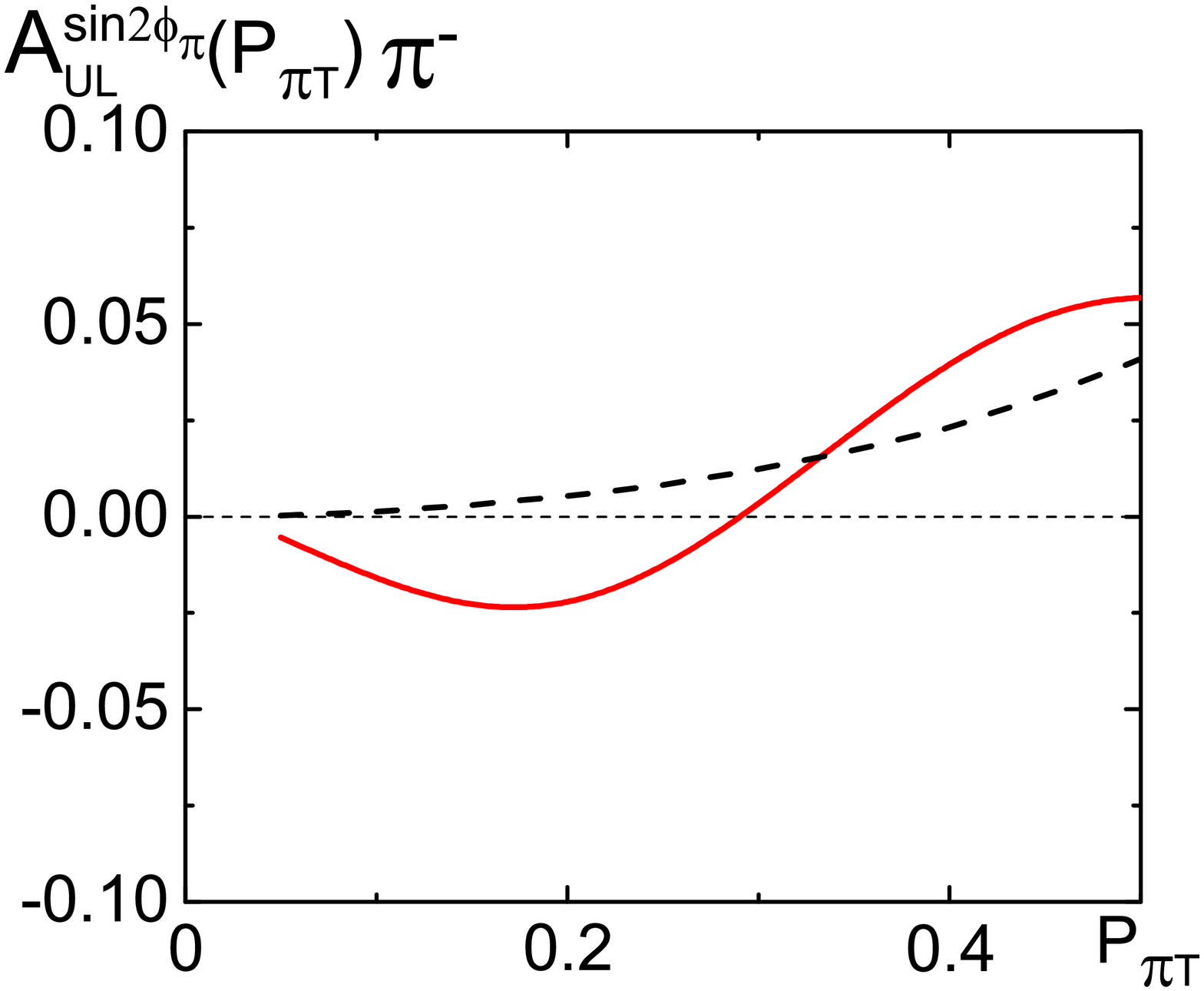}
  \includegraphics[width=0.329\columnwidth]{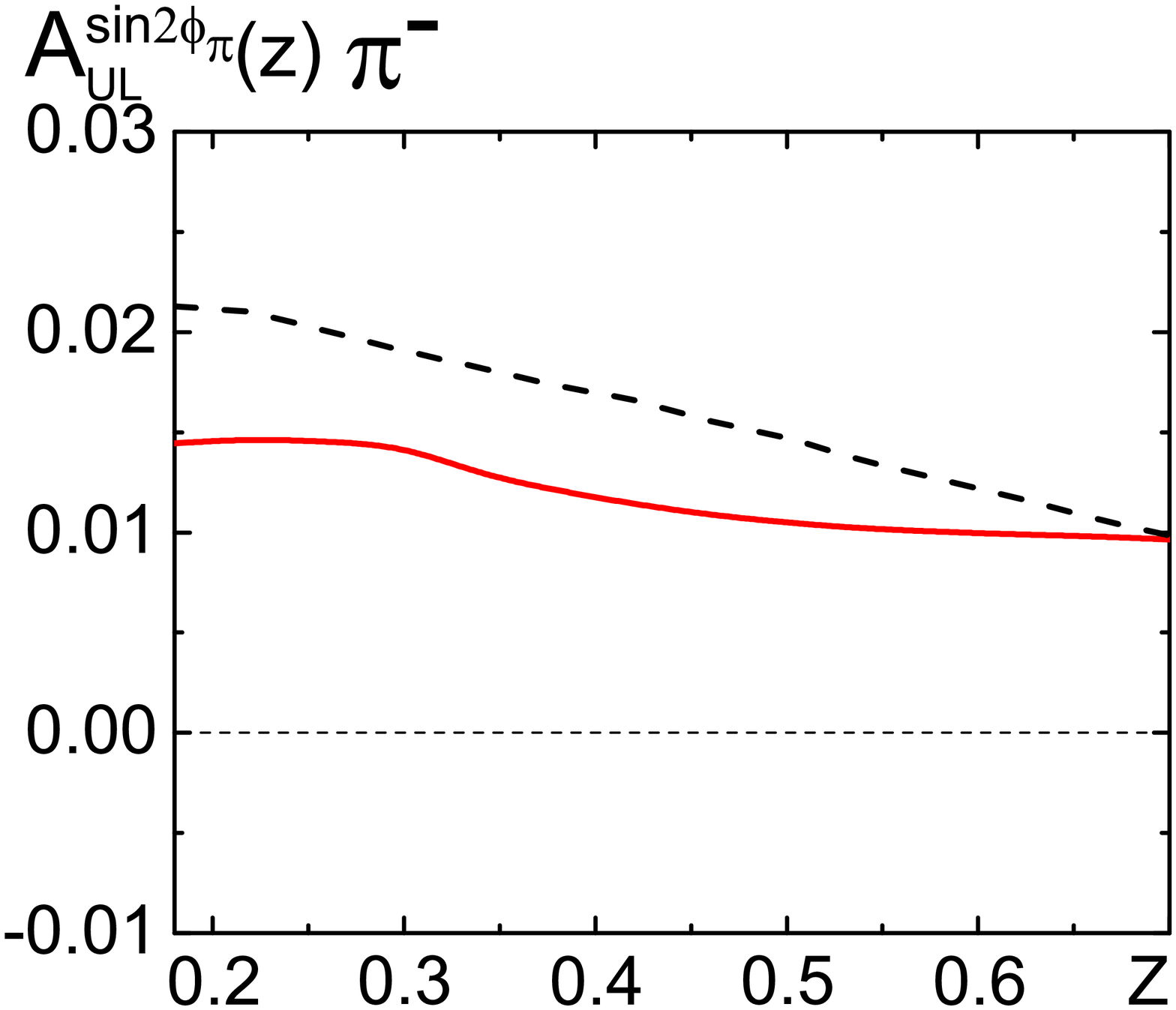}
  \includegraphics[width=0.329\columnwidth]{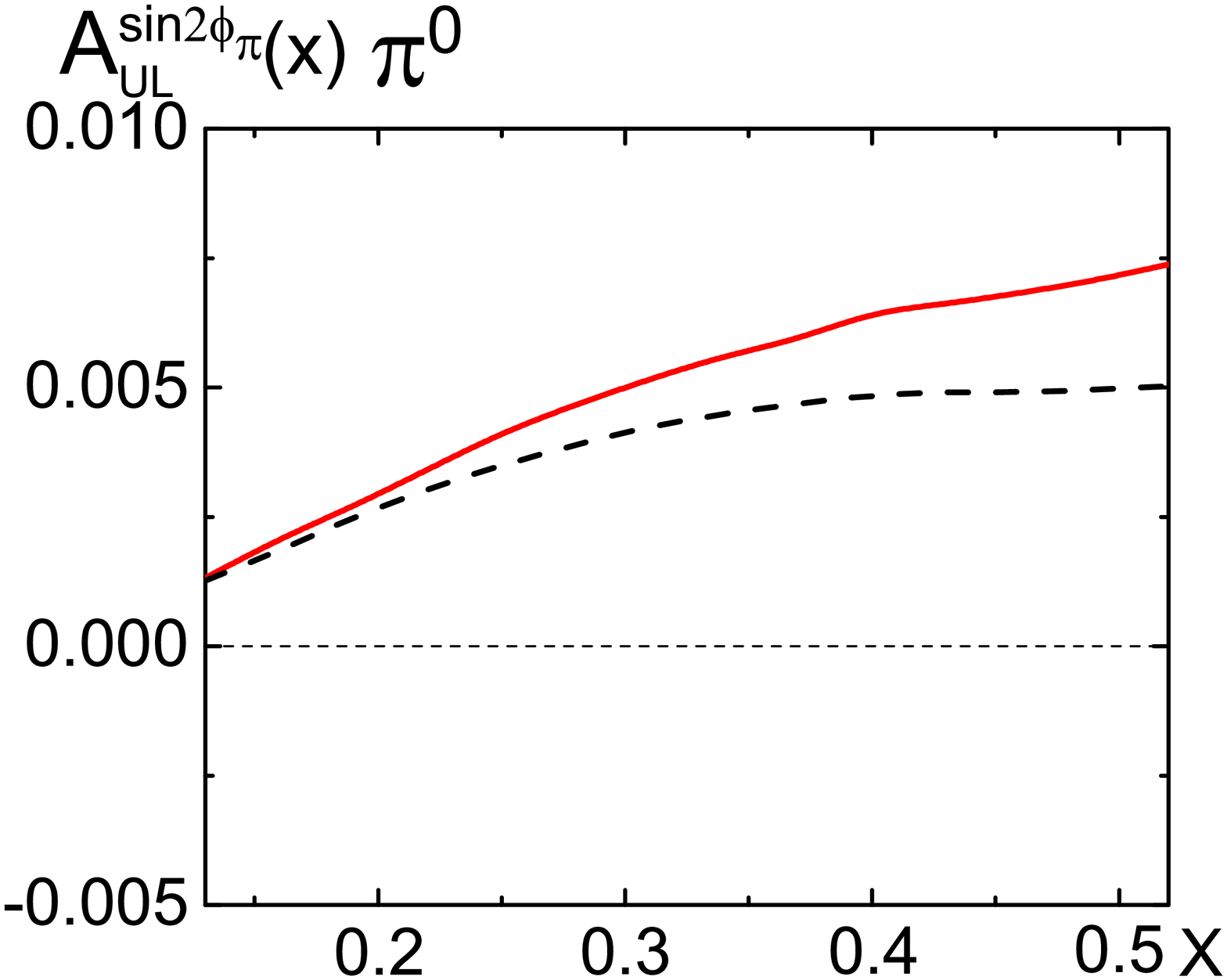}
  \includegraphics[width=0.329\columnwidth]{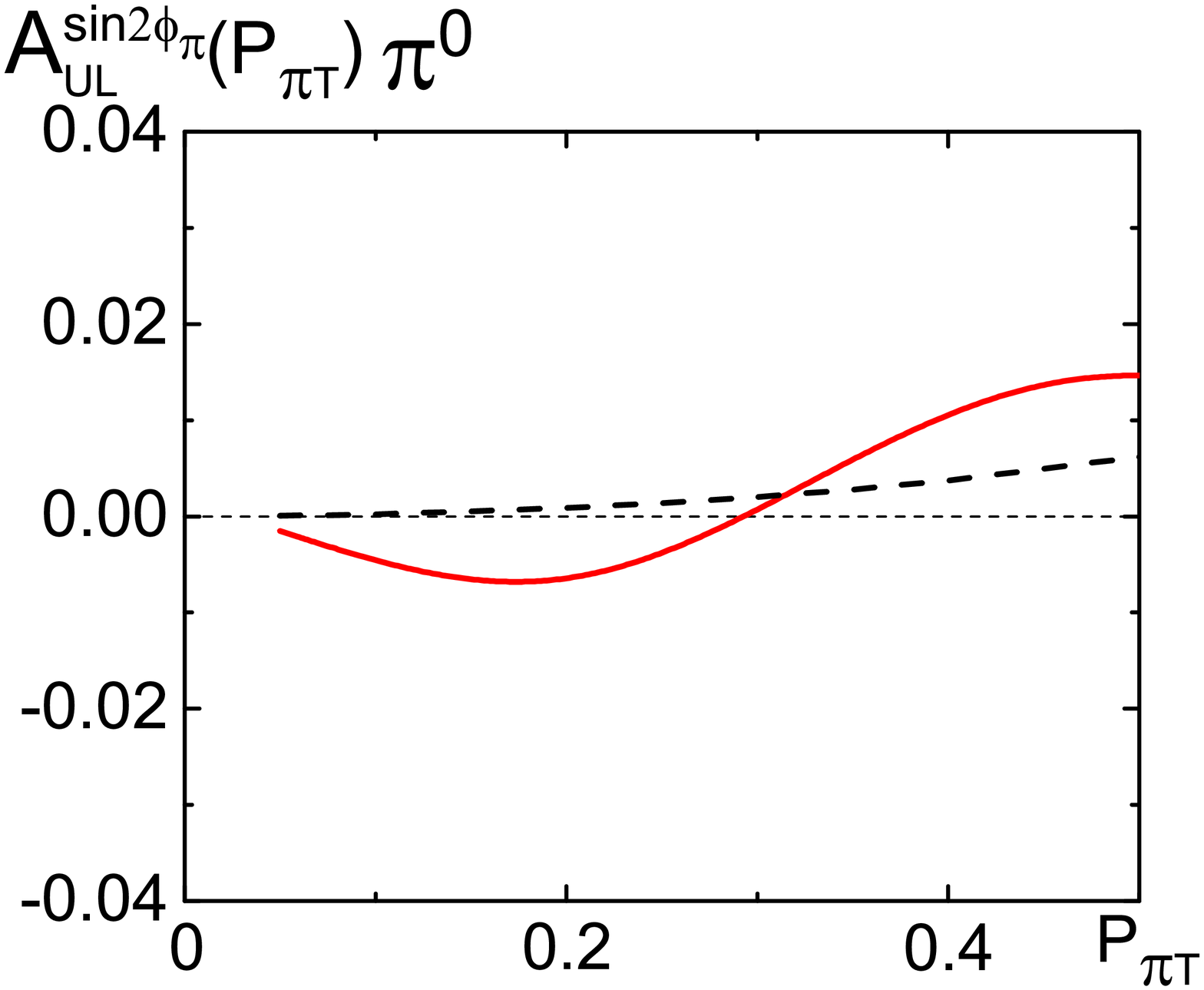}
  \includegraphics[width=0.329\columnwidth]{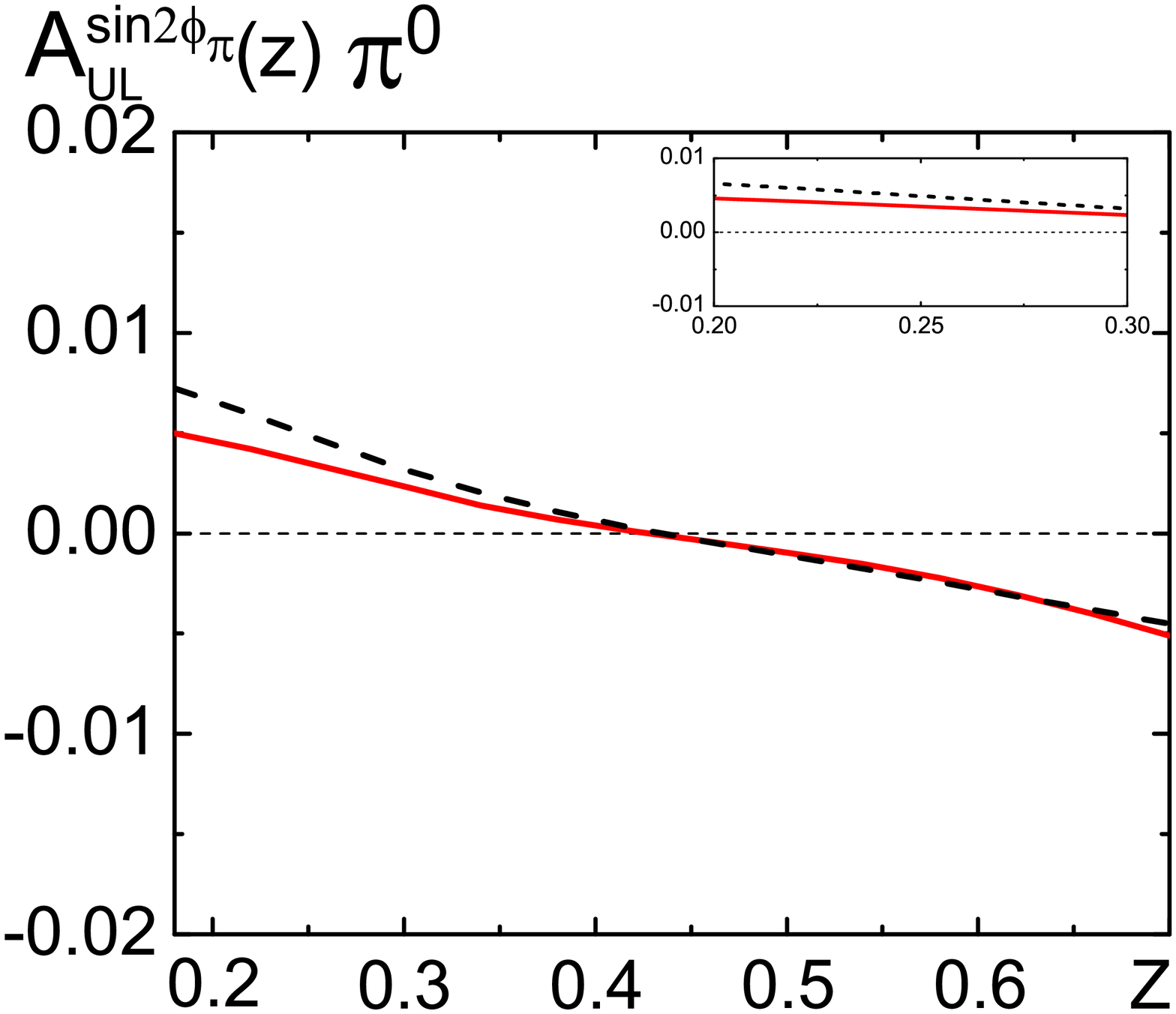}
  \caption{The $A_{UL}^{\sin2\phi_h}$ asymmetry as functions of $x$, $P_{\pi T}$ and $z$ for $\pi^+$ (upper panel), $\pi^-$ (centre panel) and $\pi^0$ (lower panel) productions at CLAS12 kinematics.}
  \label{fig3:CLAS12}
\end{figure}

The estimated asymmetry of $\pi^+$ (upper panel), $\pi^-$ (central panel) and $\pi^0$ (lower panel) productions vs $x$, $P_{\pi T}$ and $z$ are plotted in Fig.~\ref{fig3:CLAS12} respectively. From Fig.~\ref{fig3:CLAS12}, we can conclude that the size of the $\sin2\phi_h$ azimuthal asymmetries at CLAS12 is larger than that at CLAS.
Thus, it has a greater opportunity to measure the  $\sin2\phi_h$ asymmetry at CLAS12.
Another observation is that the $P_{\pi T}$-dependent asymmetries from the EIKV parametrization and the BDPRS parametrization are somewhat different. These need to be further tested by the future CLAS12 data on the proton target.

Finally, we would like to comment on the uncertainty from the theoretical aspects.
First, one should be cautious on the validation of the TMD factorization in the kinematical region of HERMES and CLAS where $Q^2$ is not so large. In our calculation we consider the region $P_T<0.5$ GeV. This should be tested by more precise data in the future. Secondly, The WW-approximation is applied for $h_{1L}^\perp$, the contribution beyond the WW-approximation may also contribute and has not considered in our calculation.
Thirdly, we apply the leading-order result for the hard coefficients, since those for $h_{1L}^\perp$ has not been established. Further studies are need to clarify these points.

\section{Conclusion}
\label{Sec.conclusion}

In this work, we applied the TMD factorization formalism to study the $\sin2\phi_h$ asymmetry in SIDIS process at the kinematical configuration of HERMES, CLAS and CLAS12.
The asymmetry provides access to the convolution of the TMD distribution $h_{1L}^{\perp}$, which describes transversely polarized quarks in the longitudinally polarized nucleon, and the Collins function of the pion.
We took into account the TMD evolution effects for these TMDs, and we adopted two approaches for the TMD evolution for comparison. One is the EIKV approach, the other is the BDPRS approach.
Their main difference is the treatment on the nonperturbative part of evolution, while the perturbative part in the two approaches are the same and was kept at NLL accuracy in this work.
As the nonperturbative part associated with $h_{1L}^{\perp}$ is still unknown, we assumed that it has the same form as that of the unpolarized distribution.
The hard coefficients associated with the corresponding collinear functions in the TMD evolution formalism are kept at the leading-order accuracy.
We also adopted the WW-type approximations for the longitudinal transversity $h_{1L}^{\perp(1)}(x,\mu)$ of the nucleon and applied the parametrization for the twist-3 fragmentation function $H_{1}^{\perp(3)}(z,z,\mu)$ to estimate the asymmetries.

The numerical calculations showed that our results generally agree with the data from HERMES and CLAS within the uncertainty bands, except for x-dependent
 asymmetry of $\pi^-$ production off the proton target, for which the sign of the asymmetry is inconsistent.
It is also found that the size of the asymmetry off the deuteron target at HERMES is sensitive to the choice of the parametrization for the nonperturbative part of evolution.
Similarly, the shape of $P_{\pi T}-$dependent asymmetries from the two parameterizations at CLAS12 are somewhat different. Future precision measurement on the asymmetry may distinguish different  parametrization on $S_{NP}$.
We also discussed the uncertainties from the theoretical aspects.
Further studies are needed on the $\sin2\phi_h$ asymmetry for a deeper understanding of nucleon structure in  three-dimensional space and the validity of WW-type approximation.

\section*{Acknowledgements}

\section{Acknowledgements}
This work is partially supported by the National Natural Science Foundation of China with grant No. 11575043.


\begin{thebibliography}{99}

\bibitem{Sivers:1989cc}
  D.~W.~Sivers,
  Phys.\ Rev.\ D {\bf 41}, 83 (1990).

\bibitem{Sivers:1990fh}
  D.~W.~Sivers,
  Phys.\ Rev.\ D {\bf 43}, 261 (1991).

\bibitem{Anselmino:1994tv}
M.~Anselmino, M.~Boglione and F.~Murgia,
Phys. Lett. B \textbf{362}, 164-172 (1995)
[arXiv:hep-ph/9503290 [hep-ph]].

\bibitem{Anselmino:1998yz}
M.~Anselmino and F.~Murgia,
Phys. Lett. B \textbf{442}, 470-478 (1998)
[arXiv:hep-ph/9808426 [hep-ph]].

\bibitem{Brodsky:2002cx}
S.~J.~Brodsky, D.~S.~Hwang and I.~Schmidt,
Phys. Lett. B \textbf{530}, 99-107 (2002)
[arXiv:hep-ph/0201296 [hep-ph]].

\bibitem{Brodsky:2002rv}
S.~J.~Brodsky, D.~S.~Hwang and I.~Schmidt,
Nucl. Phys. B \textbf{642}, 344-356 (2002)
[arXiv:hep-ph/0206259 [hep-ph]].

\bibitem{Collins:2002kn}
J.~C.~Collins,
Phys. Lett. B \textbf{536}, 43-48 (2002)
[arXiv:hep-ph/0204004 [hep-ph]].

\bibitem{Ji:2002aa}
X.~d.~Ji and F.~Yuan,
Phys. Lett. B \textbf{543}, 66-72 (2002)
[arXiv:hep-ph/0206057 [hep-ph]].

\bibitem{Belitsky:2002sm}
A.~V.~Belitsky, X.~Ji and F.~Yuan,
Nucl. Phys. B \textbf{656}, 165-198 (2003)
[arXiv:hep-ph/0208038 [hep-ph]].

\bibitem{Boer:2003cm}
D.~Boer, P.~J.~Mulders and F.~Pijlman,
Nucl. Phys. B \textbf{667}, 201-241 (2003)
[arXiv:hep-ph/0303034 [hep-ph]].

\bibitem{Ji:2006ub}
X.~Ji, J.~W.~Qiu, W.~Vogelsang and F.~Yuan,
Phys. Rev. Lett. \textbf{97}, 082002 (2006)
[arXiv:hep-ph/0602239 [hep-ph]].

\bibitem{Ji:2006vf}
X.~Ji, J.~w.~Qiu, W.~Vogelsang and F.~Yuan,
Phys. Rev. D \textbf{73}, 094017 (2006)
[arXiv:hep-ph/0604023 [hep-ph]].

\bibitem{Ji:2006br}
X.~Ji, J.~W.~Qiu, W.~Vogelsang and F.~Yuan,
Phys. Lett. B \textbf{638}, 178-186 (2006)
[arXiv:hep-ph/0604128 [hep-ph]].

\bibitem{Bacchetta:2008xw}
A.~Bacchetta, D.~Boer, M.~Diehl and P.~J.~Mulders,
JHEP \textbf{08}, 023 (2008)
[arXiv:0803.0227 [hep-ph]].

\bibitem{Mulders:1995dh}
  P.~J.~Mulders and R.~D.~Tangerman,
  Nucl.\ Phys.\ B {\bf 461}, 197 (1996)
  Erratum: [Nucl.\ Phys.\ B {\bf 484}, 538 (1997)]
  [hep-ph/9510301].

\bibitem{Boer:1997nt}
D.~Boer and P.~J.~Mulders,
Phys. Rev. D \textbf{57}, 5780-5786 (1998)
[arXiv:hep-ph/9711485 [hep-ph]].

\bibitem{Bacchetta:2006tn}
A.~Bacchetta, M.~Diehl, K.~Goeke, A.~Metz, P.~J.~Mulders and M.~Schlegel,
JHEP \textbf{02}, 093 (2007)
[arXiv:hep-ph/0611265 [hep-ph]].

\bibitem{HERMES:2004mhh}
A.~Airapetian \textit{et al.} [HERMES],
Phys. Rev. Lett. \textbf{94}, 012002 (2005)
[arXiv:hep-ex/0408013 [hep-ex]].

\bibitem{HERMES:2009lmz}
A.~Airapetian \textit{et al.} [HERMES],
Phys. Rev. Lett. \textbf{103}, 152002 (2009)
[arXiv:0906.3918 [hep-ex]].

\bibitem{HERMES:2010mmo}
A.~Airapetian \textit{et al.} [HERMES],
Phys. Lett. B \textbf{693}, 11-16 (2010)
[arXiv:1006.4221 [hep-ex]].

\bibitem{COMPASS:2005csq}
V.~Y.~Alexakhin \textit{et al.} [COMPASS],
Phys. Rev. Lett. \textbf{94}, 202002 (2005)
[arXiv:hep-ex/0503002 [hep-ex]].

\bibitem{COMPASS:2006mkl}
E.~S.~Ageev \textit{et al.} [COMPASS],
Nucl. Phys. B \textbf{765}, 31-70 (2007)
[arXiv:hep-ex/0610068 [hep-ex]].

\bibitem{COMPASS:2010hbb}
M.~G.~Alekseev \textit{et al.} [COMPASS],
Phys. Lett. B \textbf{692}, 240-246 (2010)
[arXiv:1005.5609 [hep-ex]].

\bibitem{Mkrtchyan:2007sr}
H.~Mkrtchyan, P.~E.~Bosted, G.~S.~Adams, A.~Ahmidouch, T.~Angelescu, J.~Arrington, R.~Asaturyan, O.~K.~Baker, N.~Benmouna and C.~Bertoncini, \textit{et al.}
Phys. Lett. B \textbf{665}, 20-25 (2008)
[arXiv:0709.3020 [hep-ph]].

\bibitem{CLAS:2008nzy}
M.~Osipenko \textit{et al.} [CLAS],
Phys. Rev. D \textbf{80}, 032004 (2009)
[arXiv:0809.1153 [hep-ex]].

\bibitem{EuropeanMuon:1986ulc}
M.~Arneodo \textit{et al.} [European Muon],
Z. Phys. C \textbf{34}, 277 (1987).

\bibitem{ZEUS:2000esx}
J.~Breitweg \textit{et al.} [ZEUS],
Phys. Lett. B \textbf{481}, 199-212 (2000)
[arXiv:hep-ex/0003017 [hep-ex]].

\bibitem{Kafer:2008ud}
W.~Kafer [COMPASS],
[arXiv:0808.0114 [hep-ex]].

\bibitem{Bressan:2009eu}
A.~Bressan [COMPASS],
[arXiv:0907.5511 [hep-ex]].

\bibitem{NA10:1986fgk}
S.~Falciano \textit{et al.} [NA10],
Z. Phys. C \textbf{31}, 513 (1986).

\bibitem{NA10:1987sqk}
M.~Guanziroli \textit{et al.} [NA10],
Z. Phys. C \textbf{37}, 545 (1988).

\bibitem{NuSea:2007ult}
L.~Y.~Zhu \textit{et al.} [NuSea],
Phys. Rev. Lett. \textbf{100}, 062301 (2008)
[arXiv:0710.2344 [hep-ex]].

\bibitem{NuSea:2008ndg}
L.~Y.~Zhu \textit{et al.} [NuSea],
Phys. Rev. Lett. \textbf{102}, 182001 (2009)
[arXiv:0811.4589 [nucl-ex]].

\bibitem{Kotzinian:1994dv}
A.~Kotzinian,
Nucl. Phys. B \textbf{441}, 234-248 (1995)
[arXiv:hep-ph/9412283 [hep-ph]].

\bibitem{Kotzinian:1997wt}
  A.~M.~Kotzinian and P.~J.~Mulders,
  Phys.\ Lett.\ B {\bf 406}, 373 (1997)
  [hep-ph/9701330].

\bibitem{Collins:1992kk}
J.~C.~Collins,
Nucl. Phys. B \textbf{396}, 161-182 (1993)
[arXiv:hep-ph/9208213 [hep-ph]].


\bibitem{Airapetian:1999tv}
  A.~Airapetian {\it et al.} [HERMES Collaboration],
  Phys.\ Rev.\ Lett.\  {\bf 84}, 4047 (2000)
  [hep-ex/9910062].

\bibitem{Airapetian:2002mf}
  A.~Airapetian {\it et al.} [HERMES Collaboration],
  Phys.\ Lett.\ B {\bf 562}, 182 (2003)
  [hep-ex/0212039].

  \bibitem{Avakian:2010ae}
  H.~Avakian {\it et al.} [CLAS Collaboration],
  Phys.\ Rev.\ Lett.\  {\bf 105}, 262002 (2010)
  [arXiv:1003.4549 [hep-ex]].

\bibitem{Lu:2011pt}
Z.~Lu, B.~Q.~Ma and J.~She,
Phys. Rev. D \textbf{84}, 034010 (2011)
[arXiv:1104.5410 [hep-ph]].

\bibitem{Zhu:2011zza}
J.~Zhu and B.~Q.~Ma,
Phys. Lett. B \textbf{696}, 246-251 (2011)
[arXiv:1104.4564 [hep-ph]].

\bibitem{Boffi:2009sh}
S.~Boffi, A.~V.~Efremov, B.~Pasquini and P.~Schweitzer,
Phys. Rev. D \textbf{79}, 094012 (2009)
[arXiv:0903.1271 [hep-ph]].

\bibitem{Ma:2000ip}
B.~Q.~Ma, I.~Schmidt and J.~J.~Yang,
Phys. Rev. D \textbf{63}, 037501 (2001)
[arXiv:hep-ph/0009297 [hep-ph]].

\bibitem{Ma:2001ie}
B.~Q.~Ma, I.~Schmidt and J.~J.~Yang,
Phys. Rev. D \textbf{65}, 034010 (2002)
[arXiv:hep-ph/0110324 [hep-ph]].

\bibitem{Collins:1981uk}
  J.~C.~Collins and D.~E.~Soper,
  Nucl.\ Phys.\ B {\bf 193}, 381 (1981)
  Erratum: [Nucl.\ Phys.\ B {\bf 213}, 545 (1983)].

\bibitem{Collins:1984kg}
  J.~C.~Collins, D.~E.~Soper and G.~F.~Sterman,
  Nucl.\ Phys.\ B {\bf 250}, 199 (1985).

\bibitem{Ji:2004wu}
  X.~D.~Ji, J.~P.~Ma and F.~Yuan,
  Phys.\ Rev.\ D {\bf 71}, 034005 (2005)
  [hep-ph/0404183].

\bibitem{Ji:2004xq}
  X.~d.~Ji, J.~P.~Ma and F.~Yuan,
  Phys.\ Lett.\ B {\bf 597}, 299 (2004)
  [hep-ph/0405085].

\bibitem{Collins:2011zzd}
  J.~Collins,
  Camb.\ Monogr.\ Part.\ Phys.\ Nucl.\ Phys.\ Cosmol.\  {\bf 32}, 1 (2011).

\bibitem{Avakian:1999rr}
  H.~Avakian [HERMES Collaboration],
  Nucl.\ Phys.\ Proc.\ Suppl.\  {\bf 79}, 523 (1999).

\bibitem{Avakian:2007mv}
  H.~Avakian, A.~V.~Efremov, K.~Goeke, A.~Metz, P.~Schweitzer and T.~Teckentrup,
  Phys.\ Rev.\ D {\bf 77}, 014023 (2008)
  [arXiv:0709.3253 [hep-ph]].

  \bibitem{Airapetian:2001eg}
  A.~Airapetian {\it et al.} [HERMES Collaboration],
  Phys.\ Rev.\ D {\bf 64}, 097101 (2001)
  [hep-ex/0104005].

  \bibitem{Jawalkar:2017ube}
  S.~Jawalkar {\it et al.} [CLAS Collaboration],
  Phys.\ Lett.\ B {\bf 782}, 662 (2018)
  [arXiv:1709.10054 [nucl-ex]].

\bibitem{Boer:2008fr}
  D.~Boer,
  Nucl.\ Phys.\ B {\bf 806}, 23 (2009)
  [arXiv:0804.2408 [hep-ph]].

\bibitem{Arnold:2008kf}
  S.~Arnold, A.~Metz and M.~Schlegel,
  Phys.\ Rev.\ D {\bf 79}, 034005 (2009)
  [arXiv:0809.2262 [hep-ph]].

\bibitem{Aybat:2011zv}
  S.~M.~Aybat and T.~C.~Rogers,
  Phys.\ Rev.\ D {\bf 83}, 114042 (2011)
  [arXiv:1101.5057 [hep-ph]].

\bibitem{Collins:2012uy}
  J.~C.~Collins and T.~C.~Rogers,
  Phys.\ Rev.\ D {\bf 87}, 034018 (2013)
  [arXiv:1210.2100 [hep-ph]].

\bibitem{Echevarria:2012pw}
  M.~G.~Echevarria, A.~Idilbi, A.~Sch\"{a}fer and I.~Scimemi,
  Eur.\ Phys.\ J.\ C {\bf 73}, 2636 (2013)
  [arXiv:1208.1281 [hep-ph]].

\bibitem{Echevarria:2012js}
  M.~G.~Echevarria, A.~Idilbi and I.~Scimemi,
  Phys.\ Lett.\ B {\bf 726} (2013) 795
  [arXiv:1211.1947 [hep-ph]].


\bibitem{Pitonyak:2013dsu}
  D.~Pitonyak, M.~Schlegel and A.~Metz,
  Phys.\ Rev.\ D {\bf 89}, 054032 (2014)
  [arXiv:1310.6240 [hep-ph]].

\bibitem{Echevarria:2014xaa}
  M.~G.~Echevarria, A.~Idilbi, Z.~B.~Kang and I.~Vitev,
  Phys.\ Rev.\ D {\bf 89}, 074013 (2014)
  [arXiv:1401.5078 [hep-ph]].

\bibitem{Kang:2015msa}
  Z.~B.~Kang, A.~Prokudin, P.~Sun and F.~Yuan,
  Phys.\ Rev.\ D {\bf 93}, no. 1, 014009 (2016)
  [arXiv:1505.05589 [hep-ph]].

\bibitem{Bacchetta:2017gcc}
  A.~Bacchetta, F.~Delcarro, C.~Pisano, M.~Radici and A.~Signori,
  JHEP {\bf 1706}, 081 (2017)
  Erratum: [JHEP {\bf 1906}, 051 (2019)]
  [arXiv:1703.10157 [hep-ph]].

\bibitem{Wang:2017zym}
  X.~Wang, Z.~Lu and I.~Schmidt,
  JHEP {\bf 1708}, 137 (2017)
  [arXiv:1707.05207 [hep-ph]].

\bibitem{Wang:2018pmx}
  X.~Wang and Z.~Lu,
  Phys.\ Rev.\ D {\bf 97}, 054005 (2018)
  [arXiv:1801.00660 [hep-ph]].

\bibitem{Li:2019uhj}
  H.~Li, X.~Wang and Z.~Lu,
  Phys.\ Rev.\ D {\bf 101}, 054013 (2020)
  [arXiv:1907.07095 [hep-ph]].

\bibitem{Idilbi:2004vb}
  A.~Idilbi, X.~d.~Ji, J.~P.~Ma and F.~Yuan,
  Phys.\ Rev.\ D {\bf 70}, 074021 (2004)
  [hep-ph/0406302].

\bibitem{Collins:1999dz}
  J.~C.~Collins and F.~Hautmann,
  Phys.\ Lett.\ B {\bf 472}, 129 (2000)
  [hep-ph/9908467].

\bibitem{Davies:1984sp}
  C.~T.~H.~Davies, B.~R.~Webber and W.~J.~Stirling,
  Nucl.\ Phys.\ B {\bf 256}, 413 (1985).

\bibitem{Ellis:1997sc}
  R.~K.~Ellis, D.~A.~Ross and S.~Veseli,
  Nucl.\ Phys.\ B {\bf 503}, 309 (1997)
  [hep-ph/9704239].

\bibitem{Landry:2002ix}
  F.~Landry, R.~Brock, P.~M.~Nadolsky and C.~P.~Yuan,
  Phys.\ Rev.\ D {\bf 67}, 073016 (2003)
  [hep-ph/0212159].

\bibitem{Konychev:2005iy}
  A.~V.~Konychev and P.~M.~Nadolsky,
  Phys.\ Lett.\ B {\bf 633}, 710 (2006)
  [hep-ph/0506225].

\bibitem{Aybat:2011ge}
  S.~M.~Aybat, J.~C.~Collins, J.~W.~Qiu and T.~C.~Rogers,
  Phys.\ Rev.\ D {\bf 85}, 034043 (2012)
  [arXiv:1110.6428 [hep-ph]].

\bibitem{Kang:2011mr}
  Z.~B.~Kang, B.~W.~Xiao and F.~Yuan,
  Phys.\ Rev.\ Lett.\  {\bf 107}, 152002 (2011)
  [arXiv:1106.0266 [hep-ph]].

\bibitem{Su:2014wpa}
  P.~Sun, J.~Isaacson, C.-P.~Yuan and F.~Yuan,
  Int.\ J.\ Mod.\ Phys.\ A {\bf 33}, 1841006 (2018)
  [arXiv:1406.3073 [hep-ph]].

\bibitem{Echevarria:2014rua}
  M.~G.~Echevarria, A.~Idilbi and I.~Scimemi,
  Phys.\ Rev.\ D {\bf 90}, 014003 (2014)
  [arXiv:1402.0869 [hep-ph]].

\bibitem{Collins:2014jpa}
  J.~Collins and T.~Rogers,
  Phys.\ Rev.\ D {\bf 91}, 074020 (2015)
  [arXiv:1412.3820 [hep-ph]].

\bibitem{Collins:2016hqq}
  J.~Collins, L.~Gamberg, A.~Prokudin, T.~C.~Rogers, N.~Sato and B.~Wang,
  Phys.\ Rev.\ D {\bf 94}, 034014 (2016)
  [arXiv:1605.00671 [hep-ph]].

\bibitem{Qiu:2000ga}
  J.~w.~Qiu and X.~f.~Zhang,
  Phys.\ Rev.\ Lett.\  {\bf 86}, 2724 (2001)
  [hep-ph/0012058].

\bibitem{Nadolsky:1999kb}
  P.~M.~Nadolsky, D.~R.~Stump and C.~P.~Yuan,
  Phys.\ Rev.\ D {\bf 61}, 014003 (2000)
  Erratum: [Phys.\ Rev.\ D {\bf 64}, 059903 (2001)]
  [hep-ph/9906280].

\bibitem{Aidala:2014hva}
  C.~A.~Aidala, B.~Field, L.~P.~Gamberg and T.~C.~Rogers,
  Phys.\ Rev.\ D {\bf 89}, 094002 (2014)
  [arXiv:1401.2654 [hep-ph]].

\bibitem{Qiu:2000hf}
  J.~w.~Qiu and X.~f.~Zhang,
  Phys.\ Rev.\ D {\bf 63}, 114011 (2001)
  [hep-ph/0012348].

\bibitem{Anselmino:2012aa}
  M.~Anselmino, M.~Boglione and S.~Melis,
  Phys.\ Rev.\ D {\bf 86}, 014028 (2012)
  [arXiv:1204.1239 [hep-ph]].

  \bibitem{Anselmino2005}
  M.~Anselmino et al.,
    Phys.\ Rev.\ D {\bf 71}, 074006 (2005)
    [hep-ph/0501196].

\bibitem{Collins:2005ie}
  J.~C.~Collins, A.~V.~Efremov, K.~Goeke, S.~Menzel, A.~Metz and P.~Schweitzer,
  Phys.\ Rev.\ D {\bf 73}, 014021 (2006)
  [hep-ph/0509076].

\bibitem{Schweitzer:2010tt}
  P.~Schweitzer, T.~Teckentrup and A.~Metz,
  Phys.\ Rev.\ D {\bf 81}, 094019 (2010)
  [arXiv:1003.2190 [hep-ph]].

\bibitem{Prokudin:2015ysa}
  A.~Prokudin, P.~Sun and F.~Yuan,
  Phys.\ Lett.\ B {\bf 750}, 533 (2015)
  [arXiv:1505.05588 [hep-ph]].

\bibitem{Yuan:2009dw}
  F.~Yuan and J.~Zhou,
  Phys.\ Rev.\ Lett.\  {\bf 103}, 052001 (2009)
  [arXiv:0903.4680 [hep-ph]].

\bibitem{Boer:2011fh}
  D.~Boer {\it et al.},
  arXiv:1108.1713 [nucl-th].

  \bibitem{Diehl:2021rnj}
  S.~Diehl {\it et al.} [CLAS Collaboration],
  arXiv:2101.03544 [hep-ex].


  \bibitem{Parsamyan:2018ovx}
  B.~Parsamyan,
  PoS DIS {\bf 2017}, 259 (2018)
  [arXiv:1801.01488 [hep-ex]].

  \bibitem{Parsamyan:2018evv}
  B.~Parsamyan,
  PoS QCDEV {\bf 2017}, 042 (2018).

\bibitem{Adolph:2016vou}
  C.~Adolph {\it et al.} [COMPASS Collaboration],
  Eur.\ Phys.\ J.\ C {\bf 78}, no. 11, 952 (2018)
  Erratum: [Eur.\ Phys.\ J.\ C {\bf 80}, no. 4, 298 (2020)]
  [arXiv:1609.06062 [hep-ex]].

\bibitem{Alekseev:2010dm}
  M.~G.~Alekseev {\it et al.} [COMPASS Collaboration],
  Eur.\ Phys.\ J.\ C {\bf 70}, 39 (2010)
  [arXiv:1007.1562 [hep-ex]].

\bibitem{Lai:2010vv}
  H.~L.~Lai, M.~Guzzi, J.~Huston, Z.~Li, P.~M.~Nadolsky, J.~Pumplin and C.-P.~Yuan,
  Phys.\ Rev.\ D {\bf 82}, 074024 (2010)
  [arXiv:1007.2241 [hep-ph]].

\bibitem{deFlorian:2007aj}
  D.~de Florian, R.~Sassot and M.~Stratmann,
  Phys.\ Rev.\ D {\bf 75}, 114010 (2007)
  [hep-ph/0703242 [HEP-PH]].

\bibitem{Bastami:2018xqd}
  S.~Bastami {\it et al.},
  JHEP {\bf 1906}, 007 (2019)
  [arXiv:1807.10606 [hep-ph]].

\bibitem{Kotzinian:1995cz}
  A.~M.~Kotzinian and P.~J.~Mulders,
  Phys.\ Rev.\ D {\bf 54}, 1229 (1996)
  [hep-ph/9511420].

\bibitem{Kotzinian:2006dw}
  A.~Kotzinian, B.~Parsamyan and A.~Prokudin,
  Phys.\ Rev.\ D {\bf 73}, 114017 (2006)
  [hep-ph/0603194].

\bibitem{Metz:2008ib}
  A.~Metz, P.~Schweitzer and T.~Teckentrup,
  Phys.\ Lett.\ B {\bf 680}, 141 (2009)
  [arXiv:0810.5212 [hep-ph]].

\bibitem{Teckentrup:2009tk}
  T.~Teckentrup, A.~Metz and P.~Schweitzer,
  Mod.\ Phys.\ Lett.\ A {\bf 24}, 2950 (2009)
  [arXiv:0910.2567 [hep-ph]].

\bibitem{Tangerman:1994bb}
  R.~D.~Tangerman and P.~J.~Mulders,
  [hep-ph/9408305].

\bibitem{Zhou:2008mz}
J.~Zhou, F.~Yuan and Z.~T.~Liang,
Phys. Rev. D \textbf{79} (2009), 114022
[arXiv:0812.4484 [hep-ph]].

\bibitem{Botje:2010ay}
M.~Botje,
Comput. Phys. Commun. \textbf{182} (2011), 490-532
[arXiv:1005.1481 [hep-ph]].



\end{thebibliography}
\end{document}